\documentclass[amssymb,amsmath,aps,showpacs,floatfix,nofootinbib,showpacs,12pt]{revtex4}
\usepackage{amssymb}
\usepackage{graphicx}
\usepackage{color}
\usepackage{soul}
\usepackage{latexsym}

\newcommand{\lsim}{\lesssim}
\newcommand{\gsim}{\gtrsim}

\def\lsim{\mathrel{\raise.3ex\hbox{$<$\kern-.75em\lower1ex\hbox{$\sim$}}}}
\def\gsim{\mathrel{\raise.3ex\hbox{$>$\kern-.75em\lower1ex\hbox{$\sim$}}}}

\def\beq{\begin{equation}}
\def\eeq{\end{equation}}
\def\beqn{\begin{eqnarray}}
\def\eeqn{\end{eqnarray}}
\def\bea{\begin{eqnarray}}
\def\eea{\end{eqnarray}}
\def\be{\begin{equation}}
\def\ee{\end{equation}}
\newcommand{\fslash}[1]{{#1 \kern -0.7em/ \kern 0.1em}}

\begin{document}

\voffset 1.25cm

\title{PAMELA data and leptonically decaying dark matter}

\author{Peng-fei Yin $^{1}$, Qiang Yuan $^{2}$, Jia Liu $^{1}$, Juan Zhang $^2$, Xiao-jun Bi
$^{3,2,4}$, Shou-hua Zhu $^{1,4}$ and Xinmin Zhang $^{5}$}

\affiliation{$^{1}$ Institute of Theoretical Physics \& State Key
Laboratory of Nuclear Physics and Technology, Peking University,
Beijing 100871, P.R. China \\
$^{2}$ Key Laboratory of Particle Astrophysics, Institute of High
Energy Physics, Chinese Academy of Sciences, Beijing 100049, P. R. China \\
$^{3}$ Center for High Energy Physics,
Peking University, Beijing 100871, P.R. China \\
$^4$ Kavli Institute for Theoretical Physics China, Chinese Academy
of Science, Beijing 100190, P.R. China \\
$^5$ Theoretical Division, Institute of High Energy Physics, Chinese
Academy of Sciences, P.O. Box 918-4, Beijing 100049, P.R. China}

\date{\today}

\begin{abstract}

Recently PAMELA released their first results on the positron and
antiproton ratios. Stimulated by the new data, we studied the cosmic
ray propagation models and calculated the secondary positron and
antiproton spectra. The low energy positron ratio can be consistent
with data in the convection propagation model. Above $\sim 10$ GeV
PAMELA data shows a clear excess on the positron ratio. However, the
secondary antiproton is roughly consistent with data. The positron
excess may be a direct evidence of dark matter annihilation or
decay. We compare the positron and anti-proton spectra with data by
assuming dark matter annihilates or decays into different final
states. The PAMELA data actually excludes quark pairs being the main
final states, disfavors gauge boson final states. Only in the case
of leptonic final states the positron and anti-proton spectra can be
explained simultaneously. We also compare the decaying and
annihilating dark matter scenarios to account for the PAMELA results
and prefer to the decaying dark matter. Finally we consider a
decaying neutralino dark matter model in the frame of supersymmetry
with R-parity violation. The PAMELA data is well fitted with
neutralino mass  $600\sim 2000$ GeV and life time $\sim 10^{26}$
seconds. We also demonstrate that neutralino with mass around 2TeV
can fit PAMELA and ATIC data simultaneously.

\end{abstract}


\maketitle

\section{Introduction}

The existence of dark matter (DM) has been confirmed by many
astronomical observations, but the nature of DM is still an open
question. Many kinds of particles in theories beyond the standard
model (SM) are proposed as DM candidates \cite{Bertone:2004pz}. The
DM particles are usually stable due to the protection of some
discrete symmetry. For example, the lightest supersymmetric particle
(LSP) in the usual supersymmetric (SUSY) model is stale due to the
conservation of R-parity, and is a well motivated candidate for DM
\cite{Jungman:1995df}.

Generally there are three kinds of strategies to detect the DM
particles: the ``collider search'' at large colliders such as
large hadron collider (LHC) or international linear collider
(ILC); the ``direct detection'' to find the signal of nuclei
recoil when DM particles scatter off the detector; the ``indirect
detection'' to search for the products from DM annihilation or
decay, such as neutrinos, photons and anti-matter particles. It is
usually a big challenge for indirect detection due to the
difficulty in discriminating the signal from the astrophysical
background. Therefore precise measurements with wide energy range
and improved resolution are necessary for DM indirect detection.

PAMELA is a satellite borne experiment designed to measure the
cosmic rays (CRs) in a wide energy range with unprecedented accuracy
\cite{Casolino:2008zm}. Recently, the PAMELA collaboration released
the first data about antiprotons and positrons \cite{Adriani:2008zr,
Adriani:2008zq}. Usually it is thought that antiprotons and
positrons are produced when CRs propagate in the Milky Way and
collide with the interstellar medium (ISM). The abundance of these
secondaries can be calculated with relatively high precision.
However, the PAMELA results show an obvious excess in the faction of
$e^+/(e^+ +e^-)$ at energies above $\sim 10$ GeV. Interestingly the
excess keeps to rise up to energy $\sim 100$ GeV. On the other hand,
the spectrum of antiprotons fits the prediction quite well. These
results confirm the previous results by HEAT \cite{Barwick:1997ig}
and AMS \cite{Aguilar:2007yf} within the error bars.

The PAMELA results may provide a direct evidence of DM in the way of
``indirect detection''. However, before resorting to the exotic
physics of DM, it is necessary to go through the possible
astrophysical sources to account for this results. The
model-independent spectral shape analysis shows that there might be
most likely a primary source with $e^+e^-$ pairs required to explain
the rise of the positron fraction \cite{Serpico:2008te}. The
non-excess of the antiproton data \cite{Adriani:2008zq} also favors
a leptonic origin of the positrons. Pulsar is thought to be a good
candidate to produce only leptons, and was used to explain the
previous HEAT data \cite{Atoian:1995ux, zhangli}. The recent
analysis show that the PAMELA data can also be fitted considering
the contributions from nearby pulsars such as Geminga and B0656+14
\cite{Yuksel:2008rf,Hooper:2008kg}. It should be noted that another
interesting conclusion in Ref.\cite{Delahaye:2008ua} shows that the
uncertainties of the propagation of CRs, the production cross
section of secondary particles and the errors of electron
measurements might lead to the underestimation of the positron
fraction, and the ``excess'' is actually not an excess.

As one possibility, the contribution from DM annihilation or decay
is also widely discussed. The scenarios about DM annihilation
include (i) annihilation to SM particle pairs, like gauge boson,
quark and lepton pairs
\cite{Cirelli:2008jk,Barger:2008su,Cholis:2008hb,Cirelli:2008pk};
(ii) virtual internal bremsstrahlung process to $e^+e^-\gamma$
\cite{Bergstrom:2008gr}; (iii) annihilation to new mediating
particle pair which would decay to $e^+e^-$
\cite{ArkaniHamed:2008qn,Pospelov:2008jd,Cholis:2008qq}, etc.
However, in the DM annihilation scenario, an unnatural large ``boost
factor'' is necessary to reproduce the PAMELA positron data. Another
problem is that the non-excess of the $\bar{p}/p$ data will set
constraint on the properties of annihilating DM
\cite{Cirelli:2008pk}, which makes the model building difficult.

Considering the difficulties to explain the large positron excess by
annihilating DM, we propose to solve the problem using decaying DM
in the present work. If there exists some tiny symmetry violation,
the DM particles can decay very slowly to SM particles. The decaying
DM models are studied extensively in literatures to explain the
observational data \cite{Buchmuller:2007ui,Bertone:2007aw,
Ibarra:2007wg,Ishiwata:2008cu,Ibarra:2008qg,Chen:2008dh,Covi:2008jy,Chen:2008yi}
or set constraints on the decay properties of DM
\cite{Berezinsky:1996pb,Kribs:1996ac,Baltz:1997ar,Takayama:2000uz,PalomaresRuiz:2007ry}.

For the self-consistence of this work, we first studied the
process of cosmic ray propagation carefully, and give realistic
propagation models to produce the background contributions to
positrons and antiprotons from CR interaction with ISM. The DM
induced positrons and antiprotons are calculated in the same
propagation models. We find that the PAMELA data can be well
fitted for a leptonically decaying DM (LDDM) model. The results
for different decay final states are discussed in detail. And a
possible model in the SUSY frame with R-parity violation is
proposed.

The paper is organized as follows. In section II, we discuss the
propagation of CRs and give the updated positron and antiproton
background estimations. In section III, we present a
model-independent approach to recover the PAMELA data and discuss
why we need the decaying DM whose decay products are mainly
leptons. In section IV, we give an example of this LDDM model and
discuss another possibility. Finally we give the summary and
discussion in section V.

\section{Propagation of Galactic cosmic rays}

In this section we will study the cosmic ray propagation model
carefully so that we can predict the positron and antiproton spectra
to compare with the PAMELA data. The charged particles propagate
diffusively in the Galaxy due to the scattering with random magnetic
field\cite{Gaisser:1990vg}. The interactions with ISM and
interstellar radiation field (ISRF) will lead to energy losses of
CRs. For heavy nuclei and unstable nuclei there are fragmentation
processes by collisions with ISM and radiactive decays respectively.
In addition, the overall convection driven by the Galactic wind and
reacceleration due to the interstellar shock will also affect the
distribution function of CRs. The propagation equation can be
written as\cite{Strong:1998pw} \beqn \frac{\partial \psi}{\partial
t} &=&Q({\bf x},p)+\nabla\cdot(D_{xx}\nabla \psi-{\bf
V_c}\psi)+\frac{\partial}{\partial p}p^2D_{pp}\frac{\partial}
{\partial p}\frac{1}{p^2}\psi \nonumber \\
 &-& \frac{\partial}{\partial p}\left[\dot{p}\psi
-\frac{p}{3}(\nabla\cdot{\bf V_c}\psi)\right]-\frac{\psi}{\tau_f}-
\frac{\psi}{\tau_r}, \label{prop} \eeqn where $\psi$ is the density
of cosmic ray particles per unit momentum interval, $Q({\bf x},p)$
is the source term, $D_{xx}$ is the spatial diffusion coefficient,
${\bf V_c}$ is the convection velocity, $D_{pp}$ is the diffusion
coefficient in momentum space used to describe the reacceleration
process, $\dot{p}\equiv{\rm d}p/{\rm d}t$ is the momentum loss rate,
$\tau_f$ and $\tau_r$ are time scales for fragmentation and
radioactive decay respectively. We describe a bit more about the
relevant terms in Eq.(\ref{prop}) in the follow.

For primary particles such as the protons and some heavy nuclei, the
source function is the product of two parts: the spatial
distribution $f(\bf x)$ and energy spectrum $q(p)$. $f(\bf x)$ can
follow the distribution of possible sources of CRs, such
as the supernova remnants (SNR) \cite{Strong:2004td}.
The injection spectrum
$q(p)\propto p^{-\gamma}$ is usually assumed to be a power law or
broken power law function with respect to momentum $p$. For
secondary particles the source function is given according to the
distributions of primary CRs and ISM
\begin{equation}
Q({\bf x},p)=\beta c\psi_p({\bf x},p)[\sigma_{\rm H}(p)n_{\rm H}
({\bf x})+\sigma_{\rm He}(p)n_{\rm He}({\bf x})], \label{secsource}
\end{equation}
where $\psi_p({\bf x},p)$ is the density of primary CRs, $\beta c$
is the velocity of injection CRs, $\sigma_{\rm H}$ and $\sigma_{\rm
He}$ are the cross sections for the secondary particles
from the progenitors of H and He targets, $n_{\rm H}$ and $n_{\rm
He}$ are the interstellar Hydrogen and Helium number densities,
respectively.

The spatial diffusion is regarded as isotropic and described using a
rigidity dependent function
\begin{equation}
D_{xx}=\beta D_0\left(\frac{\rho}{\rho_0}\right)^{\delta}.
\end{equation}
The reacceleration is described by the diffusion in momentum space.
The momentum diffusion coefficient $D_{pp}$ relates with the spatial
diffusion coefficient $D_{xx}$ as\cite{seo94}
\begin{equation}
D_{pp}D_{xx}=\frac{4p^2v_A^2}{3\delta(4-\delta^2)(4-\delta)w},
\end{equation}
where $v_A$ is the Alfven speed, $w$ is the ratio of
magnetohydrodynamic wave energy density to the magnetic field energy
density, which characterizes the level of turbulence. $w$ can be
taken as $1$ and the reacceleration is determined by the Alfven
speed $v_A$\cite{seo94}.

The convection velocity, which corresponds to the Galactic wind, is
assumed to be cylindrically symmetric and increase linearly with the
height $z$ from the Galactic plane\cite{Strong:1998pw}. It means a
constant adiabatic energy loss of CRs. $V_c(z=0)=0$ is adopted to
avoid the discontinuity across the Galactic plane.

Finally the energy losses and fragmentations can be calculated
according to the interactions between CRs and ISM or ISRF.

For some simplified cases the propagation equation (\ref{prop}) can
be solved analytically using the Green's function
method\cite{Kamionkowski:1991nj, Baltz:1998xv,Maurin:2001sj}.
However, generally it is not easy to find the analytical solution. A
numerical method to solve this equation has been developed by Strong
and Moskalenko, known as the GALPROP model
\cite{Strong:1998pw,Moskalenko:1997gh}. In GALPROP, the realistic
astrophysical inputs such as the ISM and ISRF are adopted to
calculate the fragmentations and energy losses of CRs. The
parameters are tuned to reproduce the observational CR spectra at
Earth. It is shown that the GALPROP model can give relative good
descriptions of all kinds of CRs, including the secondaries such as
$e^+$, $\bar{p}$ and diffuse
$\gamma$-rays\cite{Strong:1998pw,Moskalenko:1997gh,
Moskalenko:2001ya,Strong:1998fr,Strong:2004de}.

\begin{table}[htb]
\caption{The propagation parameters in the DC and DR models.}
\begin{tabular}{ccccccc}
\hline \hline
  & $D_0$ & diffusion index\footnotemark[1] & $v_A$ & ${\rm d}V_c/{\rm d}z$ & $e^-$ injection\footnotemark[2]\ \ & nuclei injection\footnotemark[3] \vspace{-3mm}\\
  & ($10^{28}$ cm$^2$ s$^{-1}$) & $\delta_1/\delta_2$ & (km s$^{-1}$) & (km s$^{-1}$ kpc$^{-1}$) & $\gamma_1/\gamma_2$ & $\gamma_1/\gamma_2$ \\
\hline
  DC & $2.5$ & $0/0.55$ & --- & $6$ & $1.50/2.54$ & $2.45/2.25$ \\
  DR & $5.5$ & $0.34/0.34$ & $32$ & --- & $1.50/2.54$ & $1.94/2.42$ \\
  \hline
  \hline
\end{tabular}
\footnotetext[1]{Below/above the break rigidity $\rho_0=4$ GV.}
\footnotetext[2]{Below/above $4$ GeV.} \footnotetext[3]{The break
energy is $25$ GeV for DC model, and $15$ GeV for DR model.}
\label{tableprop}
\end{table}

\begin{figure}[!htb]
\begin{center}
\scalebox{0.6}{\includegraphics{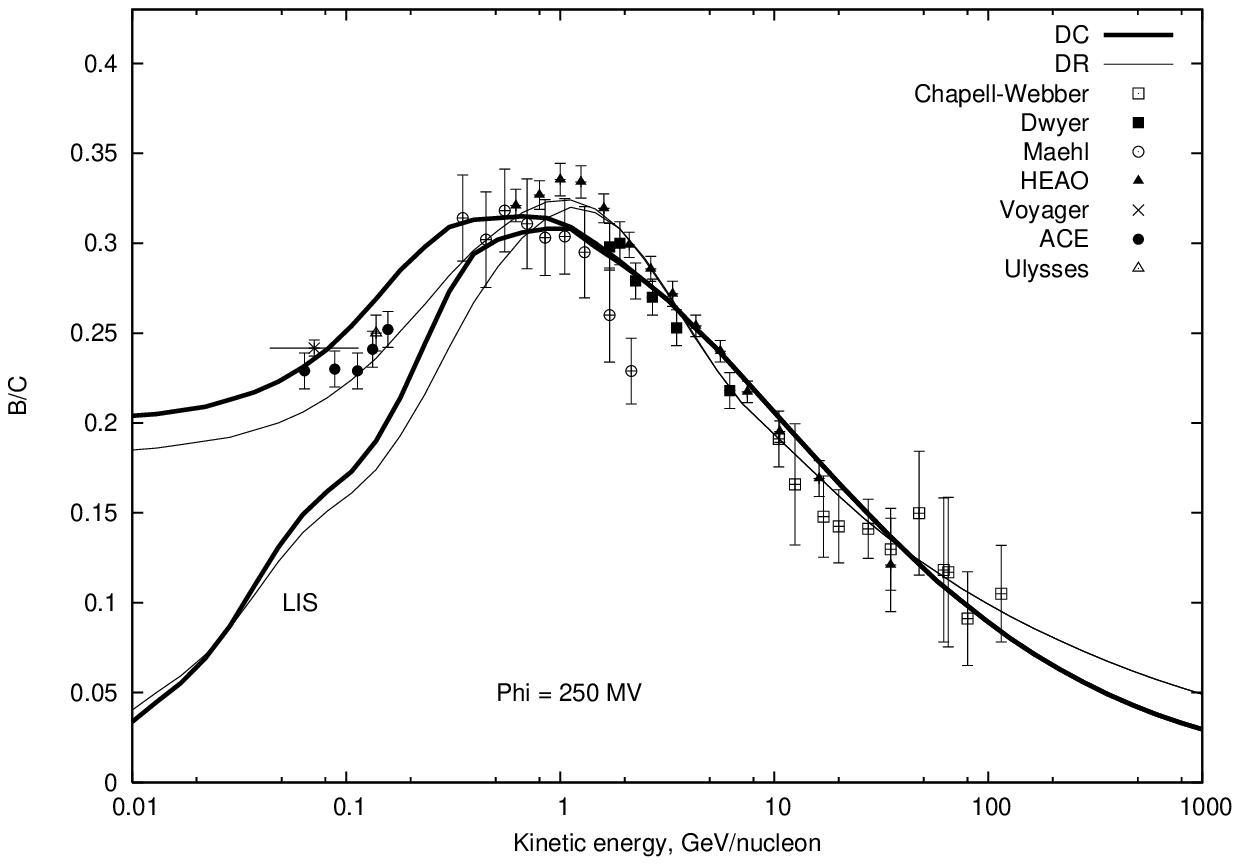}}
\scalebox{0.6}{\includegraphics{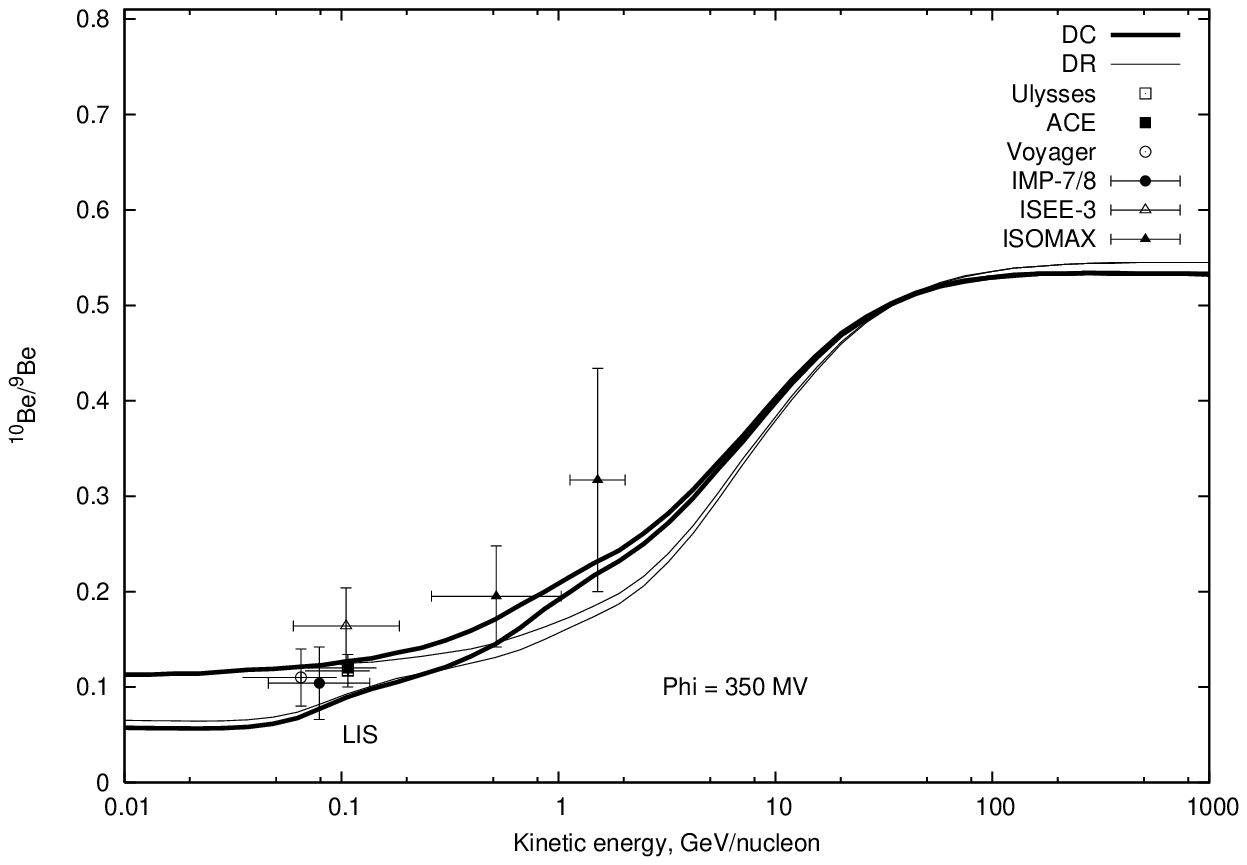}}
\scalebox{0.6}{\includegraphics{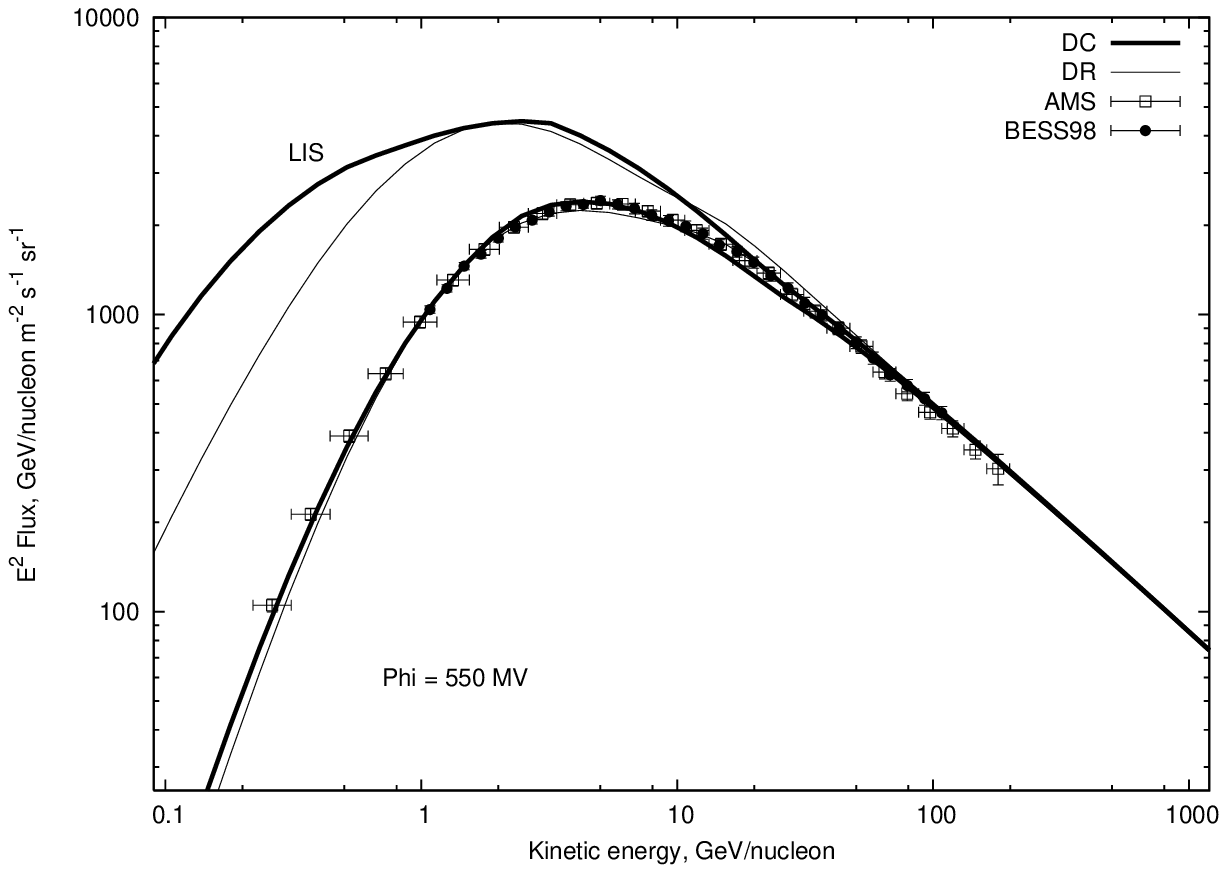}}
\scalebox{0.6}{\includegraphics{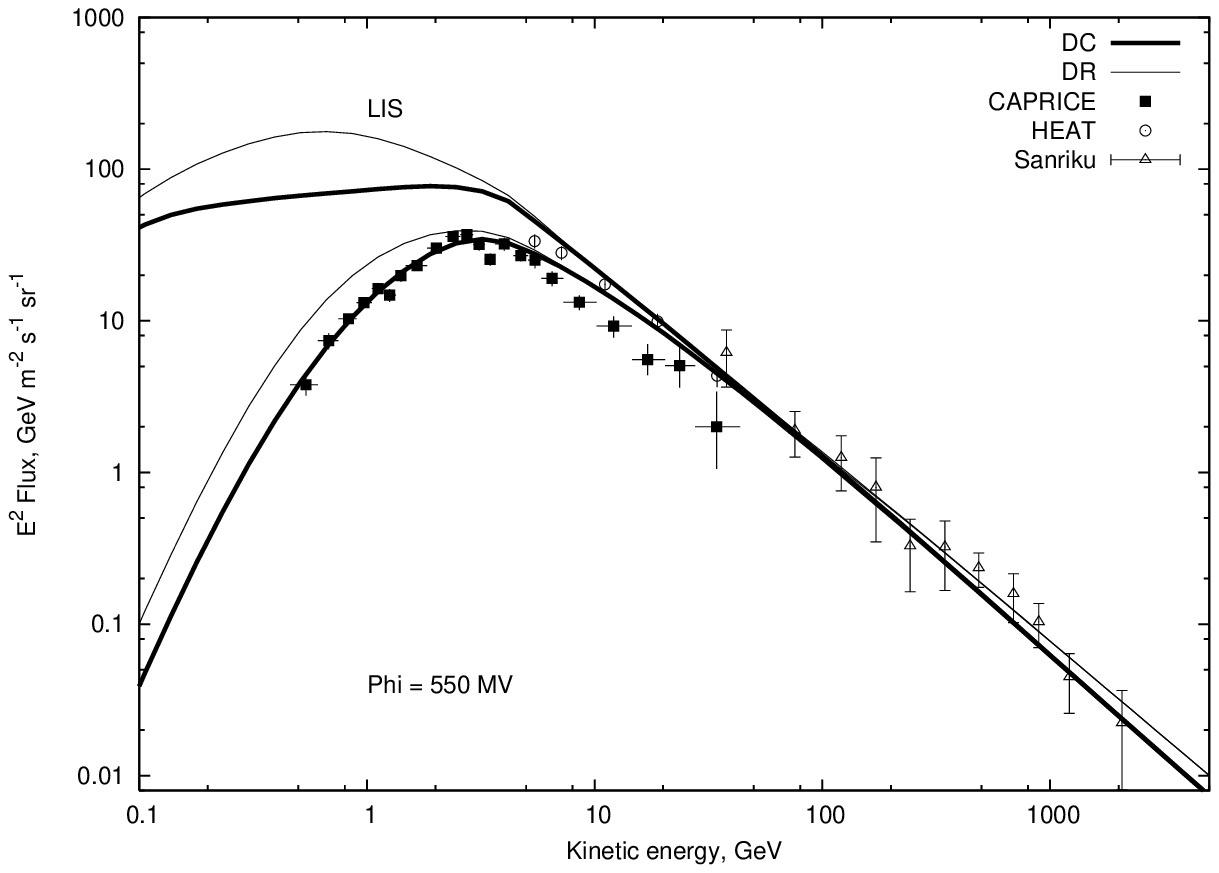}} \caption{Propagated
B/C, $^{10}$Be/$^{9}$Be, protons and electrons spectra in the DC and
DR models of GALPROP. In each panel of the figure, the thick solid
lines represent the results of DC model, while the thin solid lines
show results of DR model. For each model the LIS spectrum together
with the solar modulated one are plotted. In order to match the low
energy data, different modulation potentials are adopted as labeled
in the figure. The references of the data are, B/C:
Chapell-Webber\cite{chapwebb81}, Dwyer\cite{dwyer78},
Maehl\cite{maehl77}, HEAO\cite{heao90}, Voyager\cite{voyager99},
ACE\cite{ace00}, Ulysses\cite{ulysses96}; $^{10}$Be/$^9$Be:
Ulysses\cite{connell98}, ACE\cite{acebe}, Voyager\cite{voyager99},
IMP-7/8\cite{simpson88}, ISEE-3\cite{simpson88}, ISOMAX\cite{Hams:2004rz};
protons: BESS98\cite{Sanuki:2000wh}, AMS98\cite{Alcaraz:2000vp};
electrons: CAPRICE\cite{boezio00}, HEAT\cite{barwick98},
Sanriku\cite{Kobayashi:1999he}. } \label{creps}
\end{center}
\end{figure}

In this work we employ GALPROP models to calculate the
propagation of CRs. Two GALPROP models are adopted. One is
the diffusion $+$ convection (DC) scenario and the other is the
diffusion $+$ reacceleration (DR) model. It has been shown that the
DR model is easier to reproduce the energy dependence of
the observed B/C data\cite{Strong:1998pw}. However, the
reacceleration will produce more low energy CRs and overestimate the
low energy spectra of electrons, positrons, protons and
Helium\cite{Lionetto:2005jd}. In addition the DR model seems to
underproduce the antiprotons
\cite{Moskalenko:2001ya,Moskalenko:2002yx}. In DC model, the results
of $e^-$, $e^+$, $p$ and $\bar{p}$ are in better agreement with the
data, but the ``peak'' around $1$ GeV of the B/C data is not well
generated in the model\cite{Strong:1998pw}. Here we quote these two
models in the sense that the differences between these models are
regarded as the uncertainties of the propagation model of CRs. The
propagation parameters are listed in Table \ref{tableprop}. Other
parameters which are not included in the table are: the height of
the propagation halo $z_h=4$ kpc, the spatial distribution of
primary CRs $f(\bf x)\propto
\left(\frac{R}{R_{\odot}}\right)^{\alpha}\exp\left(
-\frac{\beta(R-R_{\odot})}{R_{\odot}}\right)\exp\left(-\frac{|z|}{z_s}
\right)$ with $\alpha=0.5$, $\beta=1.0$, $R_{\odot}=8.5$ kpc and
$z_s=0.2$ kpc\cite{strong96}.

In Fig.\ref{creps} we show the observed and calculated CR
spectra of B/C, $^{10}$Be/$^{9}$Be, protons and electrons for both
the DC and DR models. For the solar modulation of the local
interstellar (LIS) spectrum we adopt the force field
approximation\cite{gleeson68}. It is shown that these two models can
both give satisfactory descriptions of the data. We can also note
that the DR model indeed produce more low energy electrons, and a
larger solar modulation potential is needed to suppress the low energy
spectra.

\begin{figure}[!htb]
\begin{center}
\scalebox{0.6}{\includegraphics{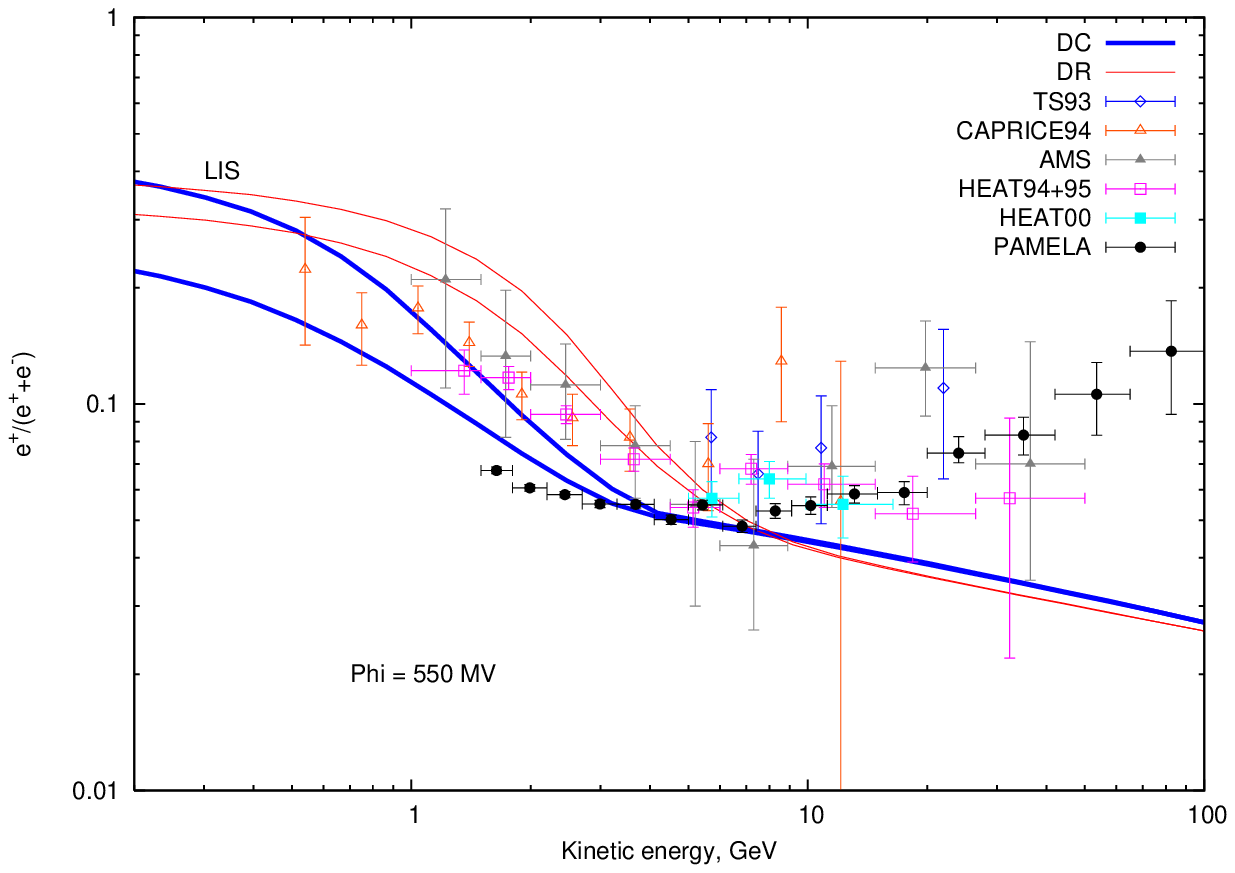}}
\scalebox{0.6}{\includegraphics{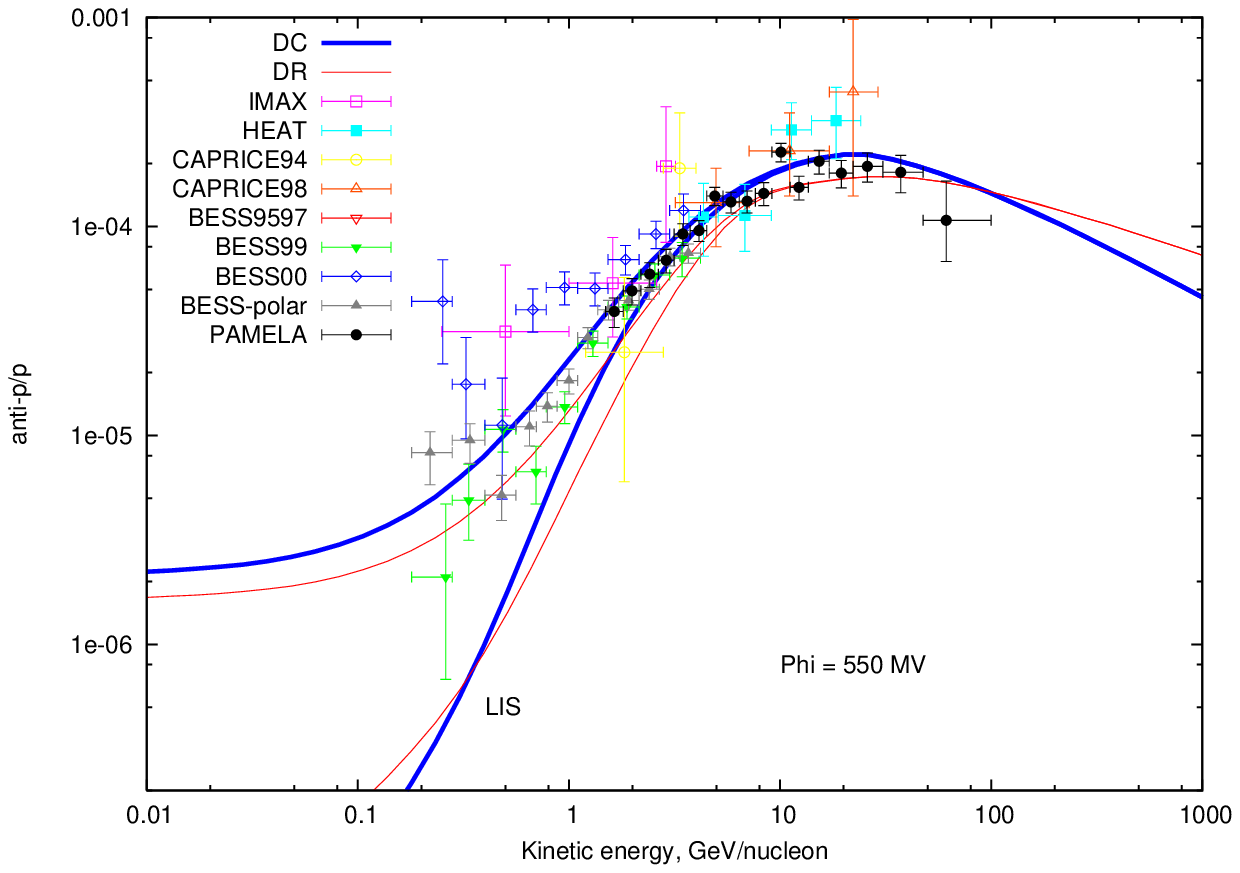}} \caption{Left: the
calculated positron fraction compared with observations; right:
$\bar{p}/p$ ratio. References of the observational data are,
positron fraction: TS93\cite{Golden:1995sq}, CAPRICE94\cite{boezio00},
AMS\cite{Aguilar:2007yf}, HEAT94+95\cite{Barwick:1997ig},
HEAT00\cite{Coutu:2001jy}, PAMELA\cite{Adriani:2008zr}; $\bar{p}/p$:
IMAX\cite{Mitchell:1996bi}, HEAT\cite{Beach:2001ub},
CAPRICE94\cite{Boezio:1997ec}, CAPRICE98\cite{Boezio:2001ac},
BESS95+97\cite{Orito:1999re}, BESS99\cite{Asaoka:2001fv},
BESS00\cite{Asaoka:2001fv}, BESS-polar\cite{Abe:2008sh},
PAMELA\cite{Adriani:2008zq}. }
\label{posipbar}
\end{center}
\end{figure}

Fig. \ref{posipbar} gives the results of positron fraction and
$\bar{p}/p$ ratio for DC and DR model respectively. We can see from
this figure that the DR model gives too many positrons at low
energies. The solar modulation does not change the results
significantly since it affects positrons and electrons
simultaneously. The charge dependent solar modulation effect might be
helpful in softening the discrepancy between the calculation and
data. The result of DC model shows a better agreement with the low
energy data. We will not focus on the low energy behavior of the
positron fraction since it might be mainly due to the solar effect.
For energies higher than several GeV, the results of the two models
are similar, and both seem to underestimate the positron fraction
compared with the HEAT \cite{Barwick:1997ig} and PAMELA data
\cite{Adriani:2008zr}. As for $\bar{p}/p$, DC model is in good
consistent with the measurements, including the recent PAMELA
data\cite{Adriani:2008zq}. The DR model shows an underproduction of
antiprotons. It means that if the DC model is correct, the excess of
positrons and non-excess of antiprotons will set strong constraints
on the properties of the source of positrons, e.g.,
\cite{Cirelli:2008pk}. On the other hand, the DR model will leave
more loose constraint.

\section{Motivation for a leptonically decaying DM model}

In the section we will adopt a model  independent approach to
constrain the DM annihilation or decay products from the PAMELA
data. We find the PAMELA data actually excludes the annihilation or
decay products being quark pairs, strongly disfavors the gauge
bosons and favors dominant leptonic final states.

In principle, there should be no difference in treating annihilation
or decay. However, there is a subtle difference for the two
scenarios. We know that the annihilation or decay depends on the DM
density in different ways. Further since what we observed on Earth
is the integrated positron or antiproton flux from the nearby
region, we may get slightly different spectra after propagation for
the annihilation or decay scenarios even the source spectra are the
same. In the section we only study the case of decay to show the
result. However, the main conclusion will be unchanged for
annihilation. In the last of the section we also show why decaying
DM is superior to annihilating DM.

The source term in Eq. \ref{prop} is given by
\begin{equation}
Q \sim
\frac{1}{\tau_{DM}}\frac{\rho(r)}{m_{DM}}\frac{dN}{dE}\mid_{decay}
\label{sourcedecay}
\end{equation}
where $\rho(r)$ is the DM density distribution in the Galaxy,
$\tau_{DM}$ is the life time of DM, $dN/dE$ is the original
positrons spectrum from each DM decay. The DM mass $m_{DM}$ is
supposed to be a free parameter while its life $\tau_{DM}$ will be
fixed by the positron fraction. In the work we take the NFW profile
for DM distribution with the local DM density
$\rho_{\odot}=0.3GeV/cm^3$.

We assume two-body final state with energy of $100$ and $300$ GeV
for either quark, lepton or gauge boson pairs ($1$ TeV is also
discussed in this case). The spectra of positron and antiproton are
simulated by PYTHIA \cite{Sjostrand:2006za} and then propagated by
adopting the DC and DR propagation models as introduced in the last
section. The life time of DM is taken to give correct positron flux.

\begin{figure}[h]
\vspace*{-.03in} \centering
\includegraphics[width=3.3in,angle=0]{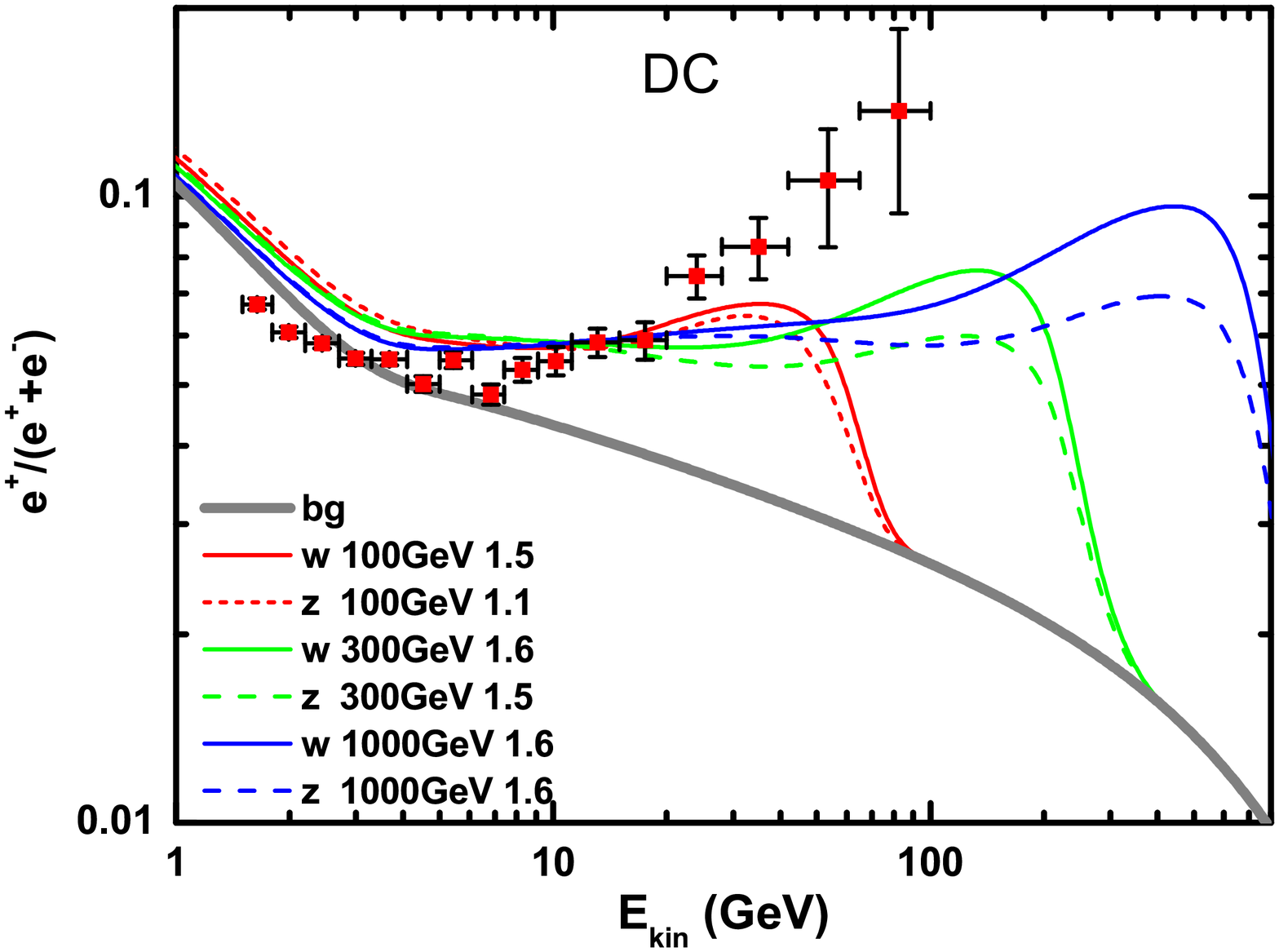}%
\includegraphics[width=3.3in,angle=0]{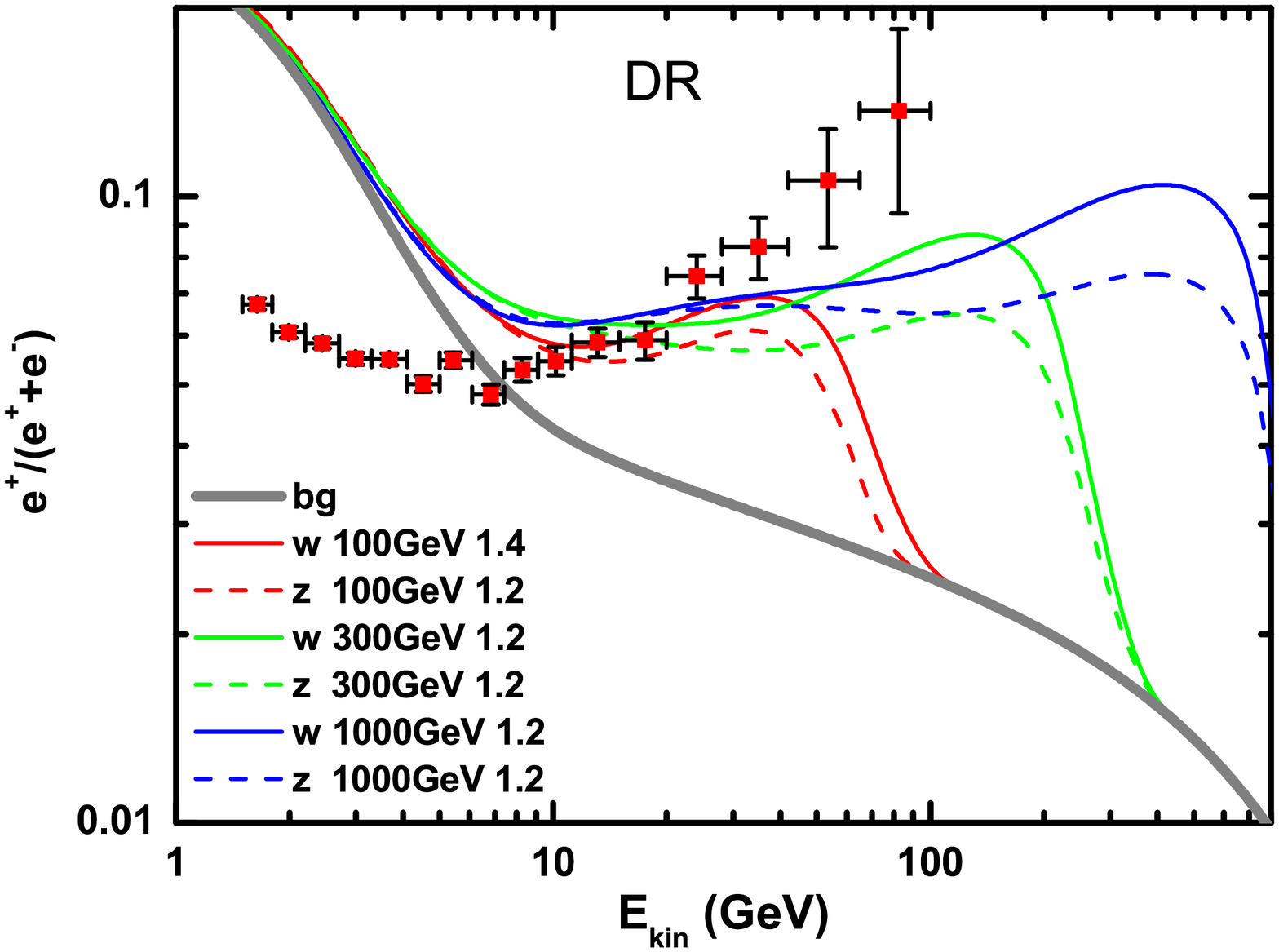}
\\
\includegraphics[width=3.3in,angle=0]{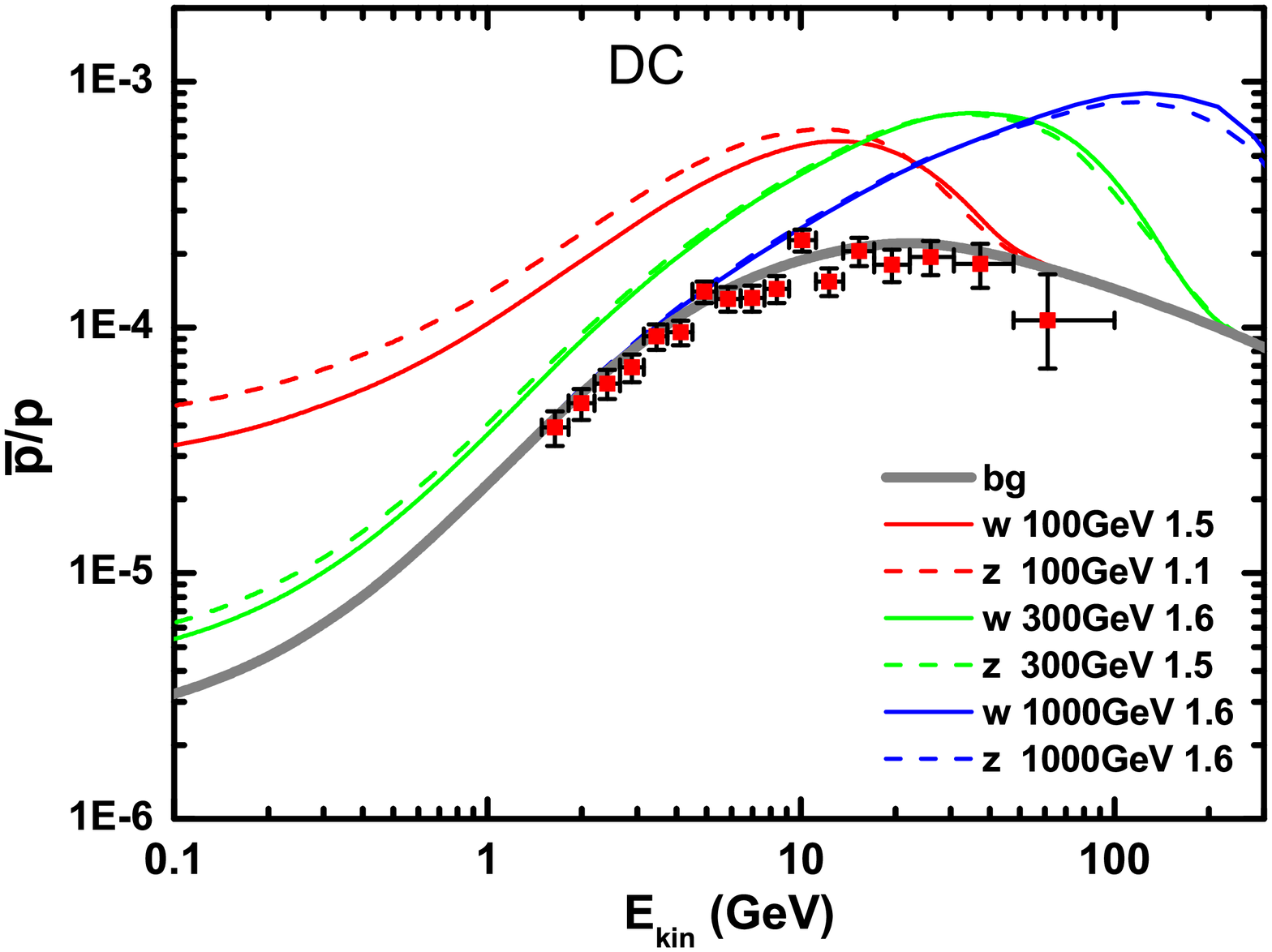}%
\includegraphics[width=3.3in,angle=0]{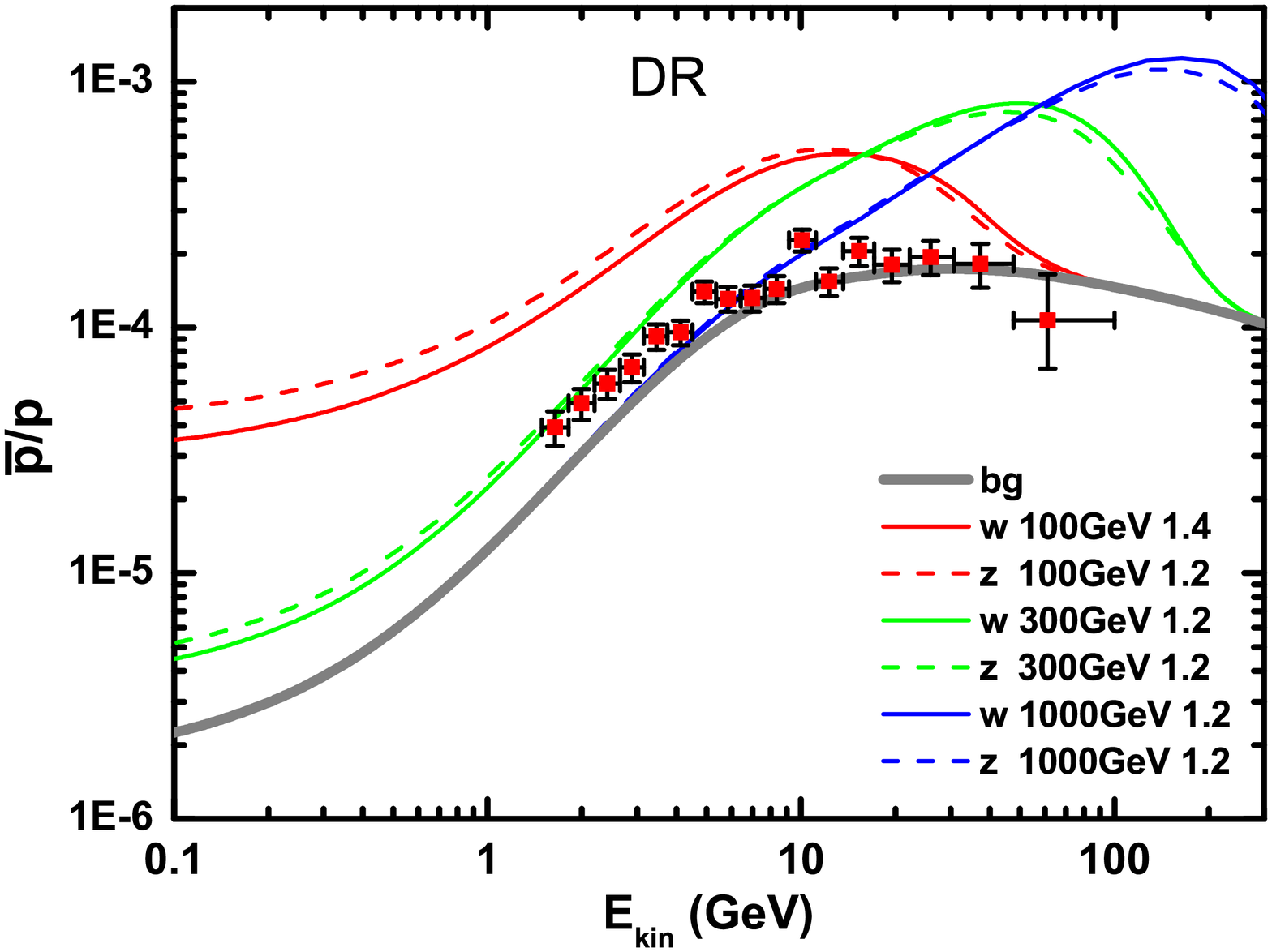}
\vspace*{-.03in} \caption{The positron and antiproton fraction, $
\frac{{\Phi _{\bar e} }}{{\Phi _{\bar e}  + \Phi _e }}$ and $
\frac{{\Phi _{\bar p} }}{{\Phi _p }}$, as function of energy from DM
decaying to gauge boson pairs. The black lines are background for
positron and anti-proton. The data points are from the preliminary
PAMELA results \cite{Adriani:2008zq}. The numbers $100$, $300$ and
$1000$ GeV refer to the energy of gauge bosons, while the numbers
3.6 and so on refer to the life time of DM in unit of $ 10^{26} s$.
In the title, ``DC'' and ``DR'' are the two propagation models
discussed in the section II.
 \vspace*{-.1in}}
\label{gaugeboson}
\end{figure}

In Fig.\ref{gaugeboson}, we show the positron and antiproton
fraction when the decay products are gauge boson pairs. We find that
the positron spectrum from gauge boson decay is usually softer than
PAMELA given even taking the gauge boson energy at $1000$ GeV.
Especially, we find the gauge boson channel is problematic for the
antiproton spectrum. They give several times larger anti-proton
fraction than the data in the two transportation models. Therefore
DM decaying to gauge bosons are strongly disfavored by the
anti-proton data. This conclusion is similar with
Ref.\cite{Cirelli:2008pk,Donato:2008jk} the authors suggest to use
extremely high energy gauge bosons of about 10TeV to interpret both
positron and antiproton spectra. In that case, only the soft tail of
positron and antiproton spectra from high energy gauge boson are
adopted. Since the endpoint of the anti-proton soft tail has energy
larger than 100 GeV it has no conflict with the present PAMELA data
that cut off at $\sim 100$ GeV. It predicts very high antiproton
flux above  $\sim 100$ GeV. The model requires very heavy DM ($\sim
10$ TeV). It should be mentioned that the
PPB-BETS\cite{Torii:2008xu} and ATIC\cite{:2008zz} experiments seem
to see electron spectrum has a significant excess above several
hundred GeV with a cutoff around 800GeV. The heavy DM with mass of
$~ 10$TeV seems not to account for this positron excess
\cite{Cirelli:2008pk}.

\begin{figure}[h]
\vspace*{-.03in} \centering
\includegraphics[width=3.3in,angle=0]{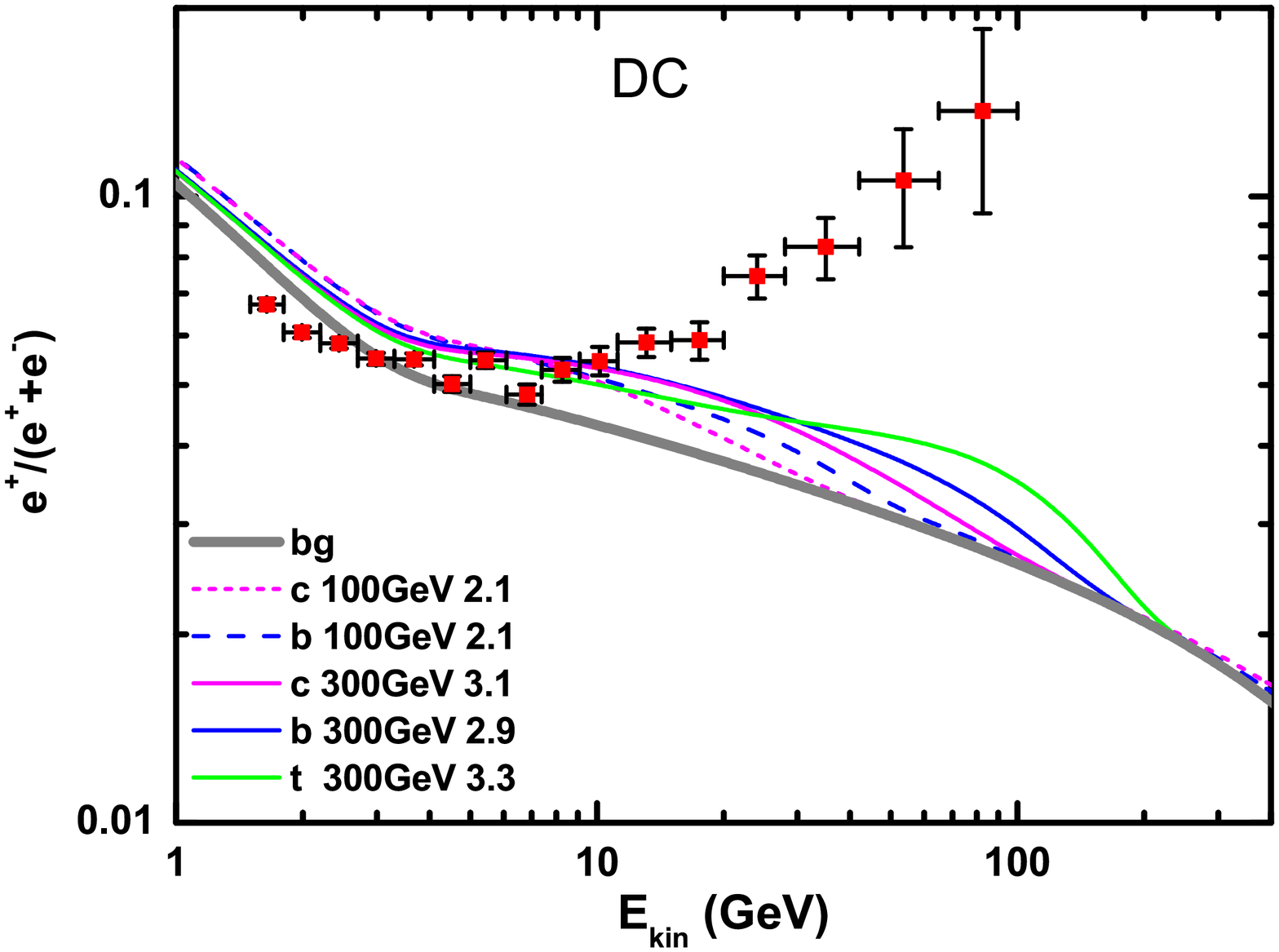}%
\includegraphics[width=3.3in,angle=0]{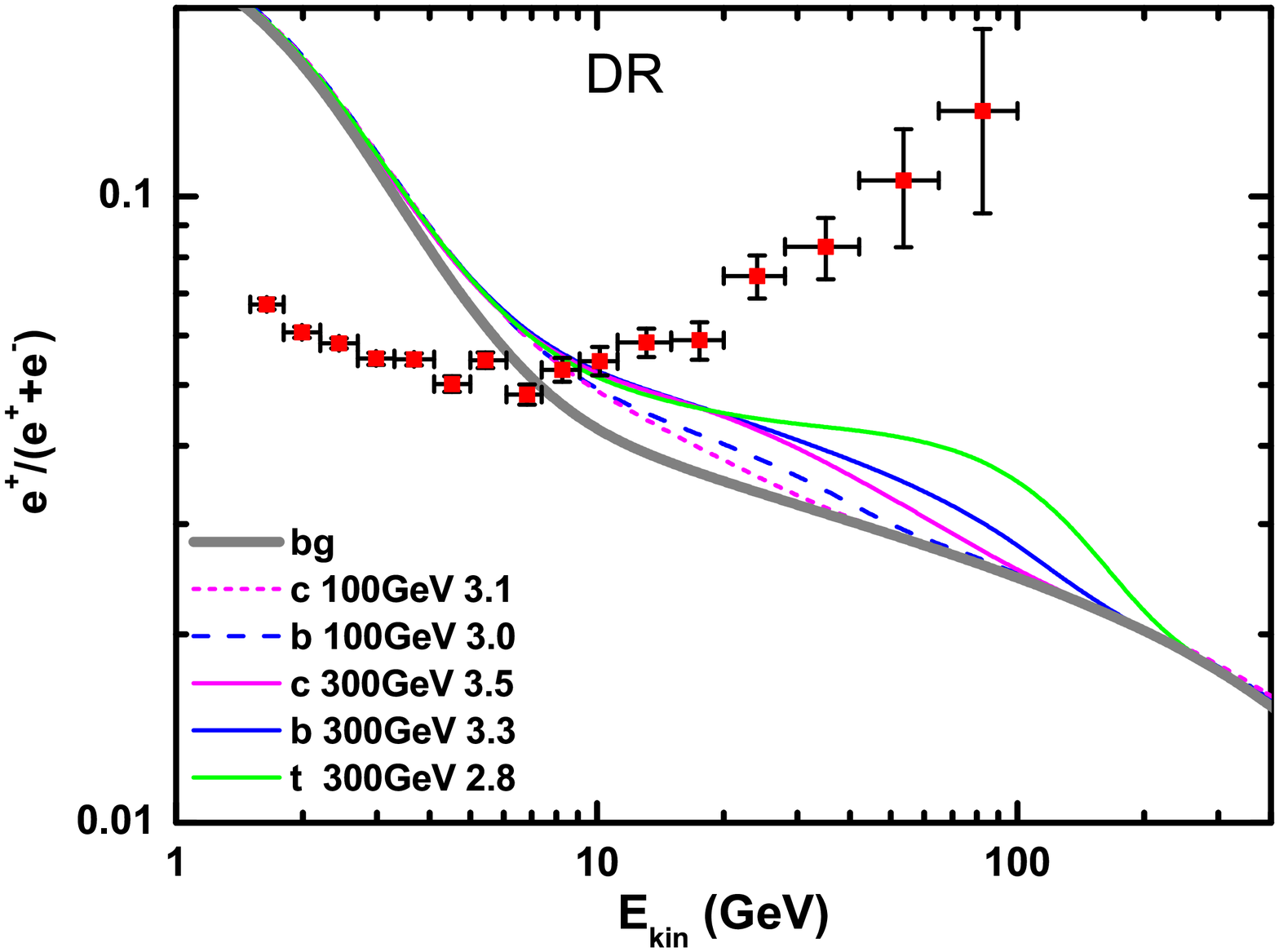}
\\
\includegraphics[width=3.3in,angle=0]{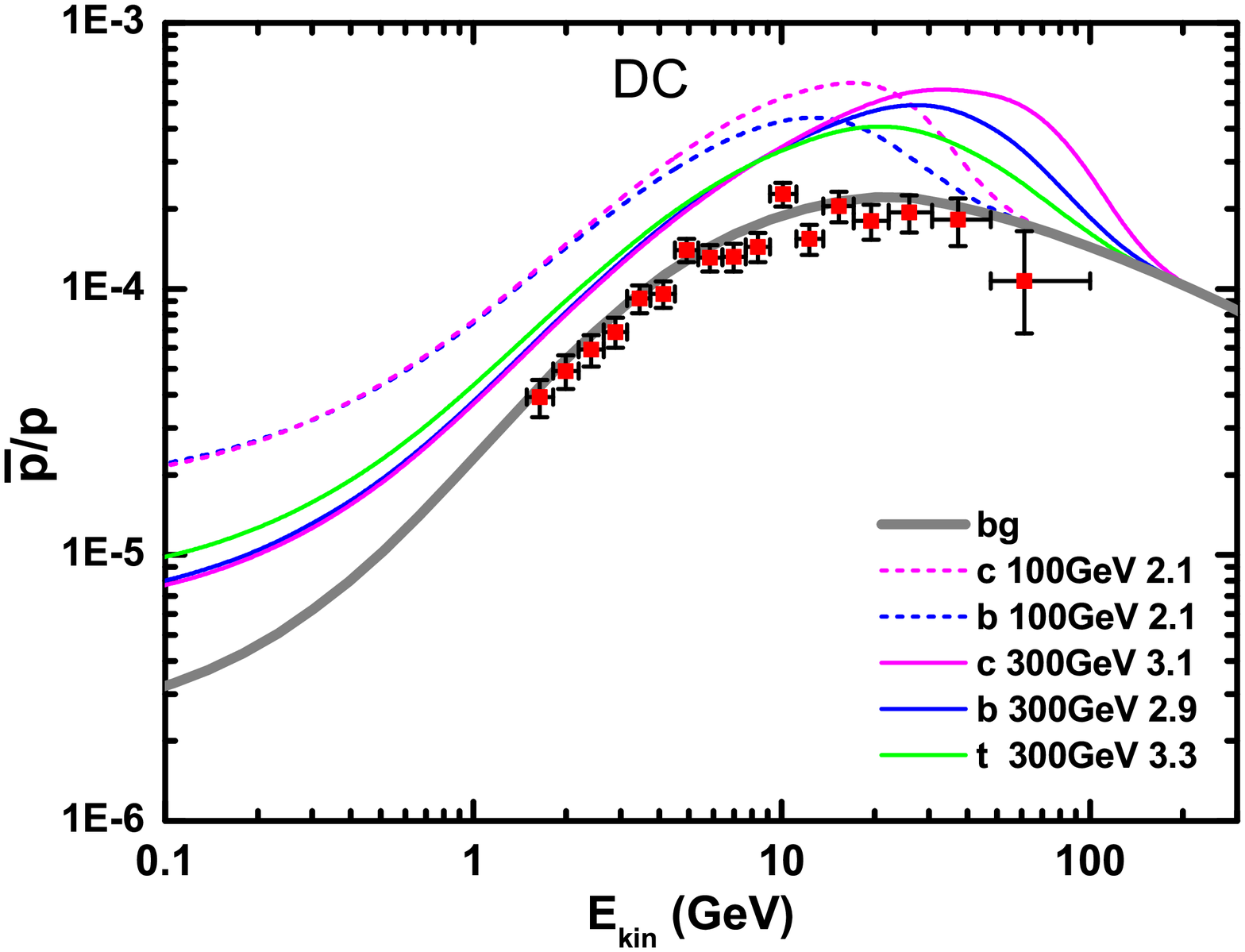}%
\includegraphics[width=3.3in,angle=0]{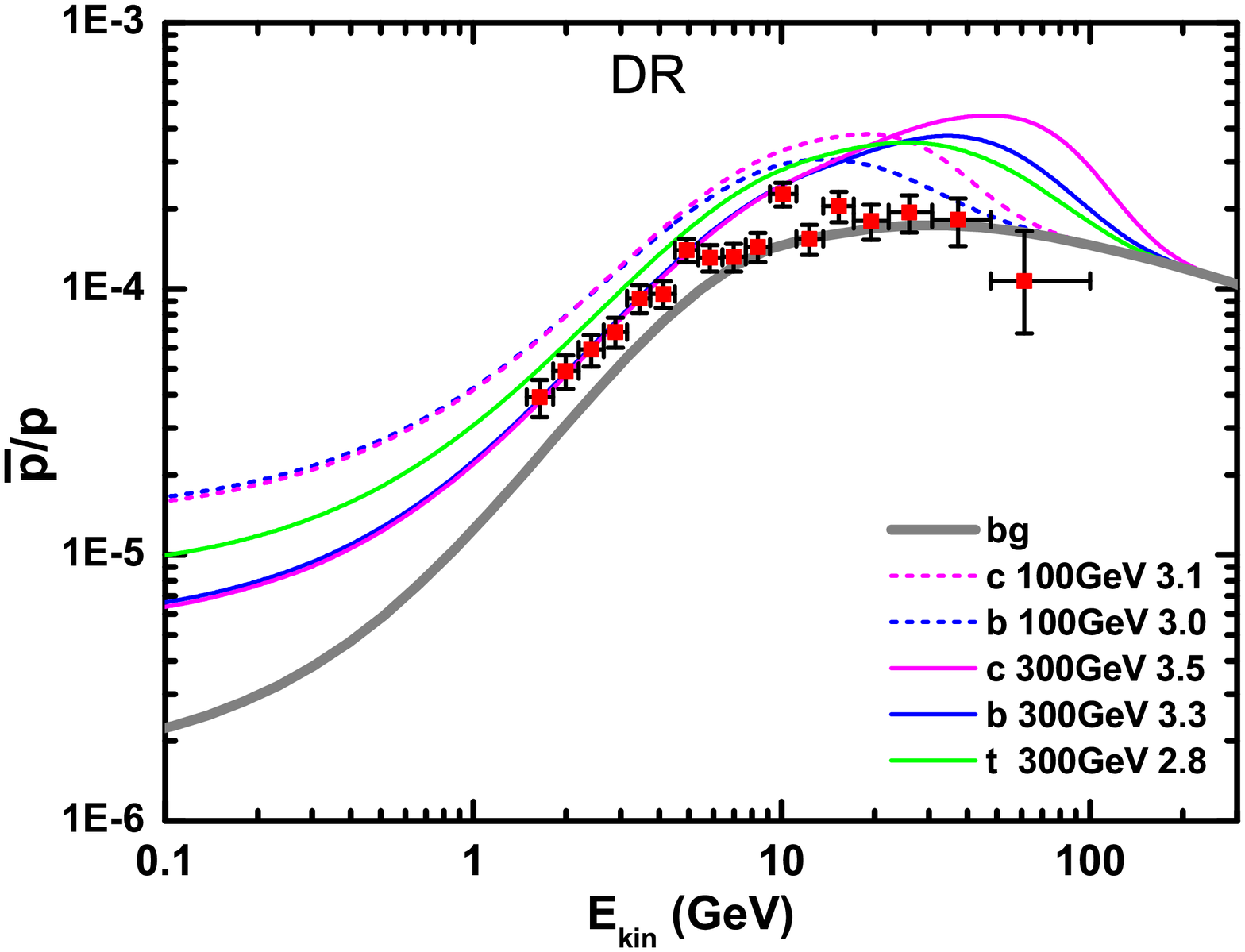}
\vspace*{-.03in} \caption{Positron and antiproton fraction
as function of energy from DM decaying to
quark pairs.
The labels are the same as that in Fig. \ref{gaugeboson}.
 \vspace*{-.1in}}
\label{quark}
\end{figure}

In Fig.\ref{quark}, we show the positron and antiproton
fraction when DM decay to quark pairs.
Positrons are produced after hadronization of quarks
via the decay of charged pions. However we find these positrons
are too soft and can not account for the excess of positrons above
$\sim 10$ GeV. Furthermore, quark hadronization produces unwanted
anti-protons as expected which are several times larger than the
experimental data.

\begin{figure}[h]
\vspace*{-.03in} \centering
\includegraphics[width=3.3in,angle=0]{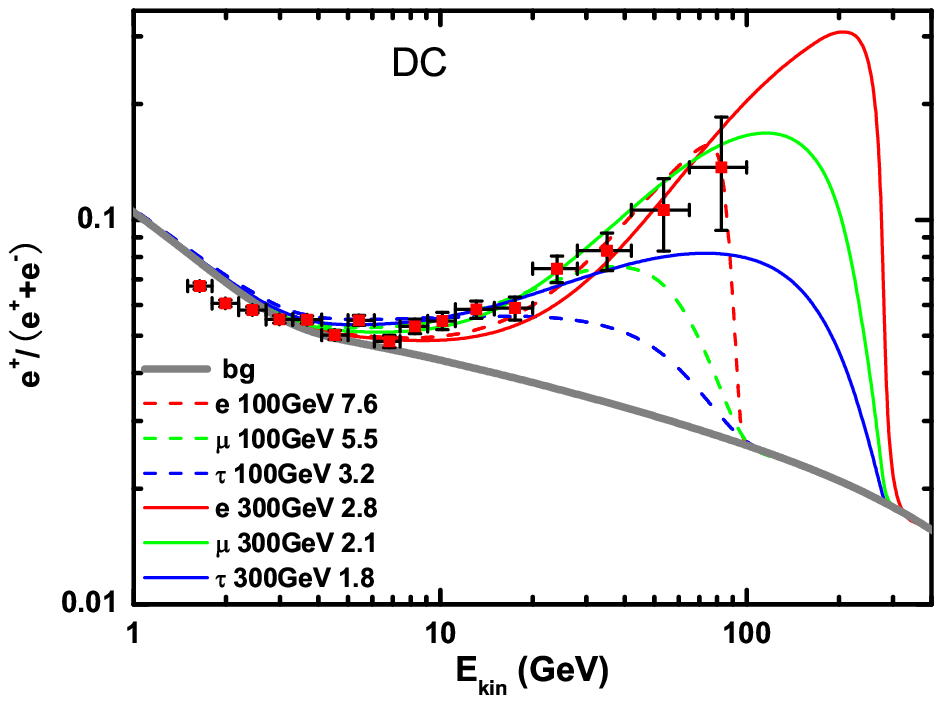}%
\includegraphics[width=3.3in,angle=0]{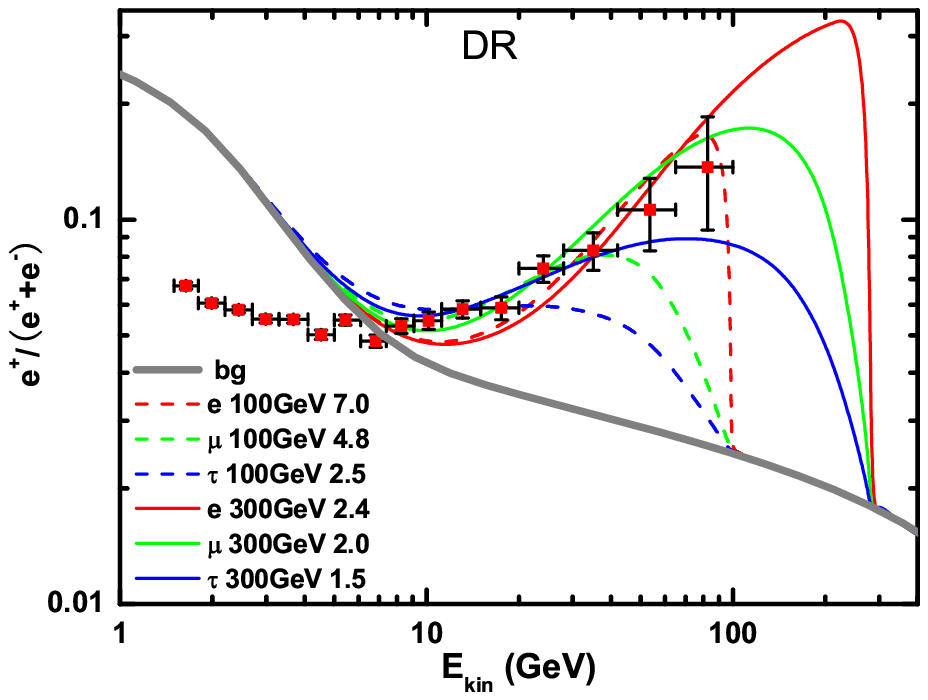}
\vspace*{-.03in} \caption{Positron and antiproton fraction
as function of energy from DM decaying to
lepton pairs.
The labels are the same as that in Fig. \ref{gaugeboson}.
 \vspace*{-.1in}}
\label{lepton}
\end{figure}

In Fig.\ref{lepton}, we show the case of leptonic final states. The
positron spectrum is easily to fit the PAMELA data in this case for
$e$ and $\mu$ final states. However, for $\tau$ final states we may
need the $\tau$ energy larger than $\sim 300$ GeV to account for the
hard positron spectrum. Certainly there is negligible influence on
the antiproton spectrum in the pure leptonic decay.

\begin{table}[htb]
\begin{tabular}{||c|c|c|c||}
\hline \hline
  & Gauge boson & Quarks & Leptons   \\
\hline
  Positron  &  $ \surd$   &  $\times$    &  $ \surd$   \\
\hline
  anti-Proton & $\times$ & $\times$   &  $ \surd$   \\
\hline \hline
\end{tabular}
\caption{The summary of three kinds of decay products to account for the PAMELA data.
 \vspace*{-.1in}}\label{table1}
\end{table}

In Table.\ref{table1} we give a summary on these three channels to
fit the PAMELA data. The gauge bosons final states with energy
around hundreds of GeV  to $\sim$TeV have problems with anti-protons
\footnote{The conclusion is drawn in the conventional propagation
model. It is possible to moderate or overcome the problem in some
special propagation models.}. However if the energy is extremely
high ($\sim 10$ TeV), the positron and anti-proton data can be
satisfied due to the energy cutoff of the PAMELA data. The quark
final states have problem in both the positron and anti-proton
spectra, that is they give too soft positron spectrum and to large
antiproton flux. The lepton final states with hundred GeV can give
hard positron spectrum and easily fit the PAMELA data very well. At
the same time they are free from upsetting the anti-proton spectrum.

Finally we will try to compare the two scenarios of annihilating and
decaying DM. We find, if the positron excess at PAMELA is indeed of
DM origin, the decaying DM is superior to annihilating DM. First,
annihilating DM has to resort to large ``boost factor'' at the order
of $\sim 10^2$ to $\sim 10^4$ to account for the large positron
flux. However, detailed analysis of boost factor from the clumpiness
of DM structures based on N-body simulation gives that the most
probable boost factor should be less than $10-20$
\cite{Lavalle:1900wn}. The same conclusion is also found through the
direct computation of the antimatter fluxes from N-body
simulation\cite{Lavalle:2008zb}. A nearby subhalo or the DM spike
around the intermediate mass black hole might be able to provide
large boost factors, however, these scenarios are found to be of
little probability\cite{Lavalle:2006vb} or suffer large
uncertainties\cite{Brun:2007tn}. Some authors use
Sommerfeld-enhancement to increase the annihilation cross-section
\cite{Cirelli:2008jk,Cirelli:2008pk,ArkaniHamed:2008qn,Pospelov:2008jd,Hisano:2004ds,MarchRussell:2008yu}
and some people adopt non-thermal DM such as wino
\cite{Grajek:2008jb}.\footnote{It is interesting to mention that DM
may be more than one-component, for example, one component is
metastable in the early Universe and the other component is stable
that it can annihilate into leptons today which can also avoid large
boost factor \cite{Fairbairn:2008fb}.} This method is usually
constrained by the data of gamma rays
\cite{Kamionkowski:2008gj,Bell:2008vx} and need further study.
Second, annihilating DM usually produce more quarks or gauge bosons
than leptonic final states. For example neutralino is more easily to
annihilate into quarks or gauge bosons than leptons. The
Kaluza-Klein DM in some universal extra dimension models may be an
exception that it can annihilate largely to leptons. However, we
should mention that Kaluza-Klein DM may still have problem on the
PAMELA antiproton data, because in its annihilation final states
quarks and gauge bosons are comparable to leptons. From Figs.
\ref{gaugeboson}, \ref{quark} and \ref{lepton} we can roughly
estimate that they may still produce too much antiprotons.

In the next section, we will consider a scenario of LDDM. Our
example is given in the minimal supersymmetric standard model with
trilinear R-parity violation term  $ LL\bar E$. The neutralino being
the lightest SUSY particle forms dark matter and is produced
thermally at early universe in the same way as stable neutralino. To
account for the PAMELA positron excess we find it has the life time
of about $\sim 10^{26}$ seconds, which is much larger than the life
of the universe.

\section{Neutralino with R-parity violation as an example of LDDM}

In this section we turn to discuss the specific decaying DM model in
supersymmetry scenario. In SUSY model, the discrete symmetry of
R-parity can be invoked to avoid dangerous baryon-number violation
terms which drive unexpected proton decay. Defined as
$R=(-1)^{2S+3B+L}$, the R-parity of a SM particle is even while
its superpartner is odd. Then the LSP particle is stable.
Since the neutralino LSP can have correct relic density via thermal
production which makes it a suitable DM
candidate.

Although R-parity symmetry is well motivated for SUSY phenomenology,
there is no reason for this symmetry to be exact. One can introduce
some R-parity violation terms in the Lagrangian which make the LSP
decay into SM particles. The general gauge invariant superpotential
of the minimal supersymmetric standard model can be written as
\begin{equation}
W = W_{MSSM}  + \lambda_{ijk} L_i L_j \bar E_k +
\lambda^{'}_{ijk}L_i Q_j \bar D_k + \lambda^{''}_{ijk}\bar U_i
\bar D_j \bar D_k + \mu^{'}_i L_i H_u \label{rvl}
\end{equation}
where i, j, k are generation indices (we neglect these indices
bellows). The $LH$ term in this superpotential mixes the lepton and
Higgs field. Because Higgs and Lepton have the same gauge quantum
numbers, one can rotate away the bilinear term by redefining these
fields. The $LL\bar E$, $LQ\bar D$ terms violate lepton number, and
the $U\bar D\bar D$ terms violate baryon number
\cite{Dreiner:1997uzBarbier:2004ez}. Then totally 45 couplings
including 9 $\lambda_{ijk}$, 27$\lambda^{'}_{ijk}$ and
9$\lambda^{''}_{ijk}$ are invoked in theory. There might also exist
many soft SUSY breaking terms which induce R-parity symmetry
breaking. These terms added more additional free parameters, and we
do not discuss them here.

The R-parity violation terms must be tiny to satisfy stringent
experiment constraints, especially the constraints from proton
decay. The tiny R-parity violation may be due to some fundamental
theories at high energy scale. For example, in SU(5) grand unified
theories (GUT) the gauge invariant terms as $f_{ijk}\bar 5^i \bar
5^j 10^k$, where $\bar 5, 10$ denote matter field representations
contain leptons and quarks, could induce R-parity violation
explicitly. $f_{ijk}$ is suppressed and gives tiny R-parity
violation couplings
\cite{Sakai:1981pk,Barbieri:1997zn,Giudice:1997wb}. To construct a
realizable theory taking account for all the experimental
constraints, more symmetries and high order operators are always
required.

In general, we should consider all R-parity violation terms in Eq.
\ref{rvl}. However, they might not appear in theory at the same
time. In some theories only baryon-number violation terms $\bar U
\bar D \bar D$ is included \cite{Tamvakis:1996xf}, while some other
theories can only predict lepton-number violation $LQ\bar D$ term
\cite{Barbieri:1997zn,Giudice:1997wb}. In a class of discrete gauge
symmetric models the baryon-number violation term should be absent
since they may induce rapid proton decay \cite{Dreiner:2005rd}. In
this work, we only consider lepton-number violation term $LL\bar E$
in order to account for the PAMELA positron excess. We assume all
the components of $ \lambda ^{ijk}$ are equal and neglect the
annihilation of neutralinos in the following discussion for
simplicity.

\begin{table}[htb]
\begin{tabular}{|c|c|c|c|c|c|c|c|}
\hline \hline  & SUSY & MC & Mass(GeV)
 & $m_0 (GeV)$
& $ m_{1/2} (GeV)$ & $ \Omega h^2$ &$ \tan \beta$
  \\
\hline
A& SPS6 & bino & 190 &  150 & 300 & $ 1.04\times 10^{ - 1}$ & 10 \\
\hline \hline & SUSY & MC & Mass(GeV)
 & $m_0 (GeV)$
& $ m_{1/2} (GeV)$ & $ \Omega h^2$ &$ \tan \beta$
  \\
  \hline
B& mSUGRA & bino & 341  & 900& 800& $ 9.62\times 10^{ - 2}$& 50 \\
\hline
C& mSUGRA & bino & 614  & 1750& 1400& $ 9.97\times 10^{ - 2}$& 50 \\
\hline
D& mSUGRA & bino & 899  & 5000& 2000& $ 1.02\times 10^{ - 1}$& 50 \\
\hline
E& mSUGRA & higgsino & 1126  & 9100& 3500& $ 1.01\times 10^{ - 1}$& 50 \\
\hline \hline & SUSY & MC & Mass(GeV)
 & $m_0 (GeV)$
& $ m_{3/2} (GeV)$ & $ \Omega h^2$ &$ \tan \beta$\\
  \hline
F& AMSB & wino & 2040  & 18000  & 640000& $ 9.15\times 10^{ - 2}$& 10 \\
  \hline
G& AMSB & wino & 2319  & 20000  & 730000& $ 1.17\times 10^{ - 1}$& 10 \\
\hline \hline
\end{tabular}
\caption{The seven benchmark points with different neutralino masses
and in different scenarios. ``MC'' means the main component of
neutralino. The other parameters we adopted are $ {\mathop{\rm sgn}}
(\mu ) =  + 1,m_t  = 172.6GeV,A_0 = 0$. The SPS6 point has
non-universal mass parameter $ M_1/1.6 = M_2 = M_3$ at GUT scale.
Suspect \cite{Djouadi:2002ze} and MicrOmega \cite{Belanger:2004yn}
are used in our
calculation.  }
\label{bench}
\end{table}

To specify the SUSY parameters we choose some benchmark points as
denoted in Tab. \ref{bench}, where LSP is neutralino with different
masses. The point A is SPS 6 in the mSUGRA-like scenario with
non-unified gaugino masses, the point B, C, D are from the Higgs
funnel region in the mSUGRA, the point E is from focus point region
in the mSUGRA, and the point F denotes a thermal wino in the AMSB
scenario. All these points satisfy the correct relic density and
other laboratory constraints. We utilize PYTHIA to produce the
positron energy spectrum from neutralino decay.

In the limit of heavy and degenerate sleptons, neutralino with
gaugino component has life time \cite{Baltz:1997ar}
\begin{equation}
\tau _{gaugino}  \sim 10^{26} s \cdot \left( {\frac{\lambda^{'}
}{{10^{ - 25} }}} \right)^{-2}  \left( {\frac{{m_\chi
}}{{1000GeV}}} \right)^{-1}\left( {\frac{{m_{\tilde f} }}{{m_\chi
}}} \right)^4
\end{equation}
where $\lambda^{'} $ is the coefficient for $ LL\bar e$ term and $
{m_{\tilde f} }$ is the sfermion mass. The neutralino with higgsino
component has life time
\begin{equation}
\tau _{higg\sin o}  \sim 10^{26} s \cdot \left( {\frac{{\tan \beta
}}{10}} \right)^{-2} \left( {\frac{\lambda^{'} }{{10^{ - 23} }}}
\right)^{-2} \left( {\frac{{m_\chi }}{{1000GeV}}} \right)^{-1}\left(
{\frac{{m_{\tilde f} }}{{m_\chi }}} \right)^4
\end{equation}

\begin{table}[htb]
\begin{tabular}{|c|c|c|c|c|c|}
\hline \hline DC &  $ \tau (10^{26} s)$
 & $\lambda^{'} (10^{ - 25} )$& DR &  $ \tau (10^{26} s)$
 & $\lambda^{'} (10^{ - 25} )$
  \\
\hline
A&  4.5 & 3.2 &  A  & 3.9 & 3.4 \\
\hline
B&  2.6 & 14.8 &  B  & 2.3 & 15.7 \\
\hline
C&  1.7 & 16.1 &  C  & 1.5 & 17.1 \\
\hline
D&  1.2 & 59.5 &  D  & 1.1 & 62.1 \\
\hline
E&  1.0 & 253.2 &  E  & 0.9 & 266.9 \\
\hline
F&  0.6 & 159.7 &  F  & 0.6 & 159.7 \\
\hline
G&  0.5 & 156.7 &  G  & 0.5 & 156.7 \\
\hline \hline
\end{tabular}
\caption{Life time $ \tau $ in unit of $10^{26} s$ and R-parity
violation parameter $\lambda^{'} $ in unit $10^{ - 25} $ for
different benchmark points. ``DC'' and ``DR'' refer to the two
propagation models we adopted.
}\label{DCDR}
\end{table}

These neutralinos generally have life time around $ 10^{26} s$ to
account for the PAMELA data. It is much longer than the life of the
universe which makes these neutralinos as valid dark matter.
However, the tiny value of $\lambda^{'}$ means we can not test the
scenario directly at LHC or ILC. In Tab. \ref{DCDR} we give the life
time and value of $\lambda^{'}$ for the different benchmark points
to account for the PAMELA positron excess.

\begin{figure}[h]
\vspace*{-.03in} \centering
\includegraphics[width=3.3in,angle=0]{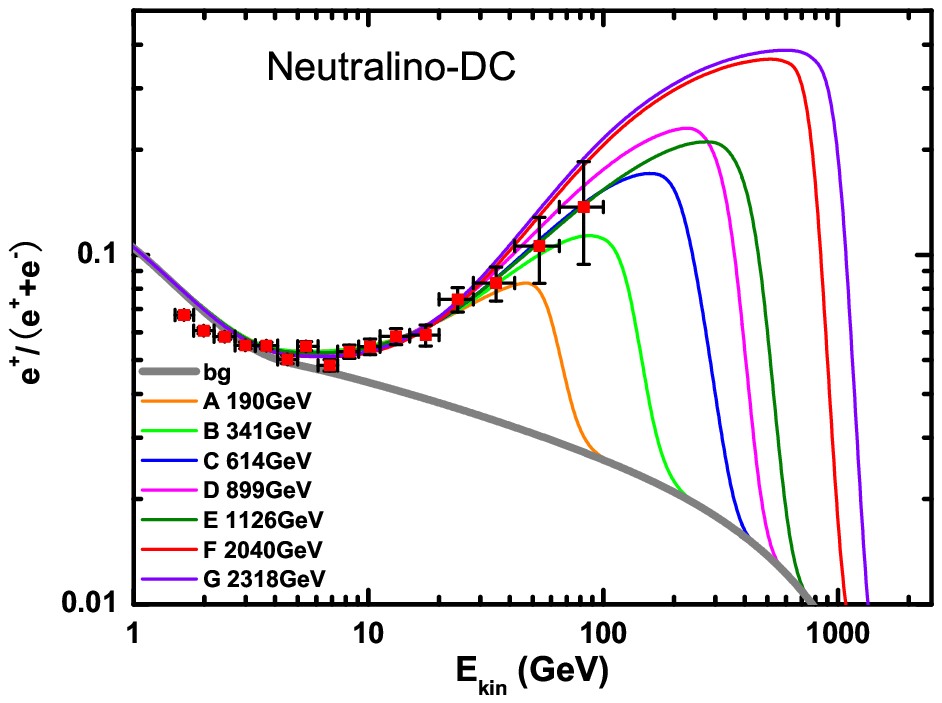}%
\includegraphics[width=3.3in,angle=0]{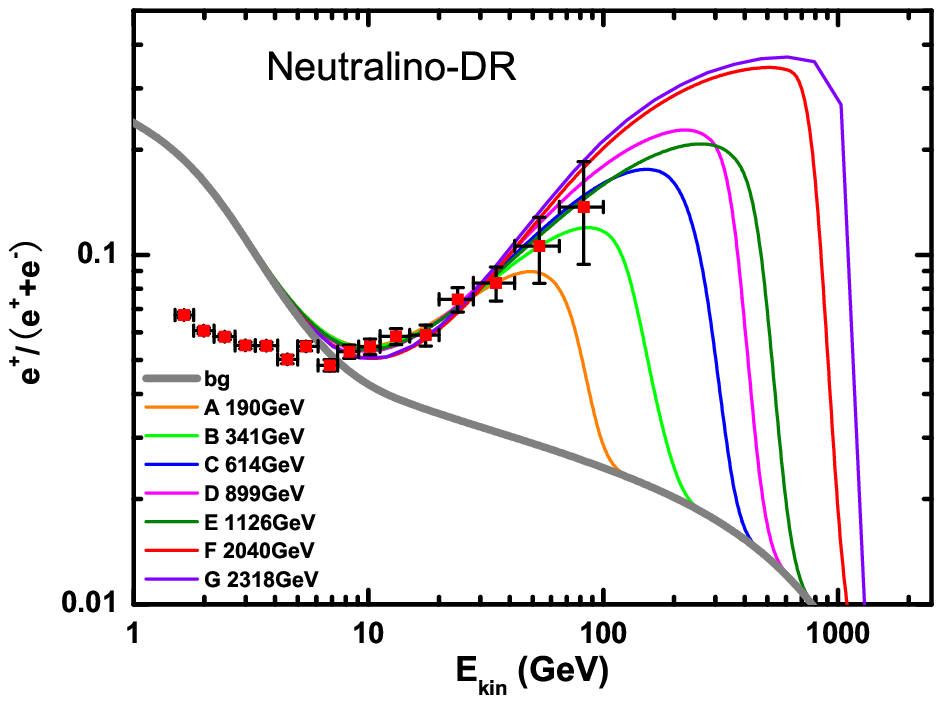}
\vspace*{-.03in} \caption{ Positron and antiproton fraction as
function of energy for different benchmark points. The neutralino
mass of each model is given, while the life time of neutralino in
each model is given in Tab.\ref{DCDR}. The neutralino mass in each
line is increasing from left to right.
}
\label{neu}
\end{figure}

In Fig. \ref{neu}, we show the positron fraction including
contribution from neutralino decay for different benchmark points.
From Fig. \ref{neu}, we can see the positrons are most probably
distributed lower than $ \frac{1}{3}m_{\tilde \chi ^0 }$. This is
due to the three-body decay and the roughly equal decay fraction of
$\nu_il^-_jl^+_k$ . The positron spectrum given here with three-body
decay is generally softer than two body final states with
monochromatic lepton. We notice that neutralino heavier than 300GeV
to about $2$ TeV can fit the PAMELA data well. If one takes account
the results from PPB-BETS and ATIC experiments, it seems to suggest
a heavy neutralinos up to $\sim$TeV if they are due to DM decay.
Another feature of this scenario is neutrino flux from neutralino
decay, but we find this neutrino flux is too small to be detected by
neutrino telescope such as Super-kamiokande, IceCube, etc while
which is well within the atmospheric neutrino bound from
Super-kamiokande(similar analysis for neutrino flux from decaying DM
can be found in Ref.\cite{Covi:2008jy,PalomaresRuiz:2007ry}).

\begin{figure}[h]
\vspace*{-.03in} \centering
\includegraphics[width=3.3in,angle=0]{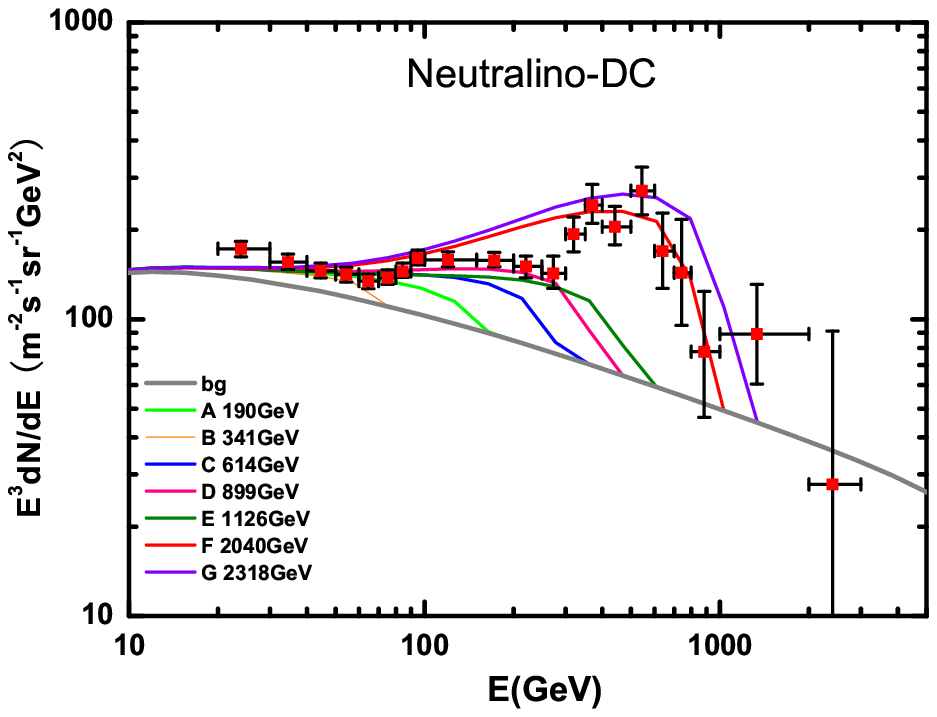}%
\includegraphics[width=3.3in,angle=0]{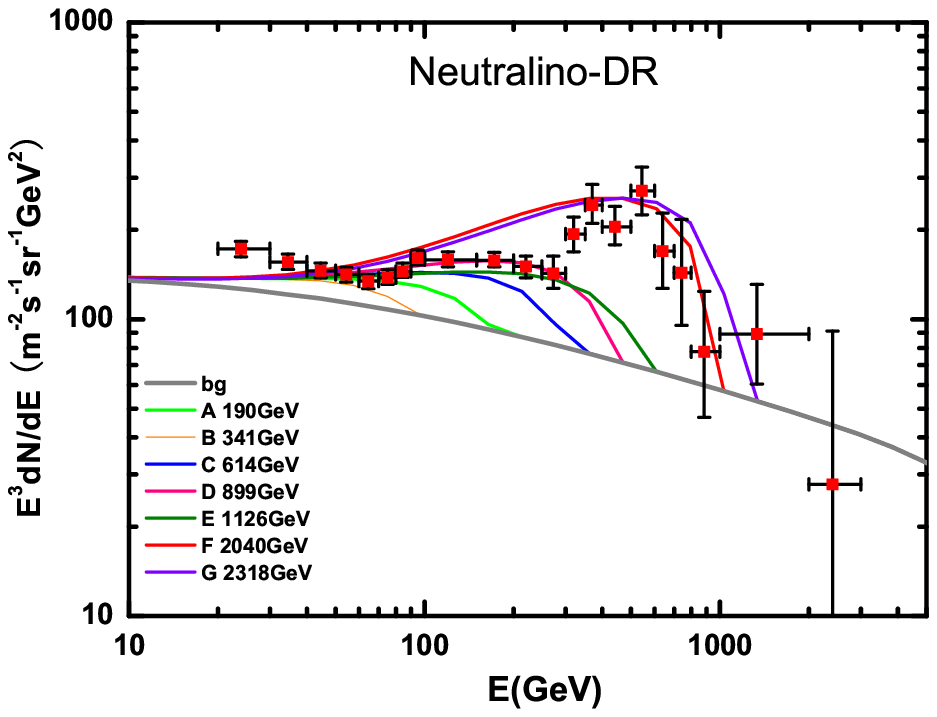}
\vspace*{-.03in} \caption{Positron and electron total flux as
function of energy for different benchmark points comparing with
ATIC data Ref.\cite{:2008zz}. The labels are the same as that in
Fig. \ref{neu}.
 \vspace*{-.1in}}
\label{ATIC}
\end{figure}

Finally we will give some comments on other scenarios of decaying
dark matter. In some SUSY scenarios, the LSP may be the neutral
particles as gravitino or sneutrino. They can also be the candidates
of DM. However, in the bilinear R-parity violation scenario, the
gravitino would mainly decay into gauge boson plus lepton
\cite{Ibarra:2008qg,Ishiwata:2008cu} which is not favored by
antiproton data from PAMELA. If the R-parity violation term is
$LL\bar E$ as we discussed above, the spectrum of decay products is
similar. But the decay of gravitino is suppressed by the Planck
scale and the R-parity violation parameter can be as large as $
\lambda^{'}_{ijk} > 10^{-14}$, which may lead to unstable NLSP decay
quickly to avoid destroying the success of big-bang nucleosynthesis
\cite{Takayama:2000uz,Buchmuller:2007ui} and be observable at
colliders \cite{Buchmuller:2007ui}.

Another possible candidate of DM in SUSY is sneutrino. The
left-handed sneutrino has been ruled out by DM direct detection,
therefore the right-handed sneutrino receives more concerns. If we
neglect all the soft breaking terms, the interactions for
right-handed neutrino superfeild $N$ are induced only from Yukawa
term $y_{ij}L_i H_u N_j$. After introducing only bilinear R-parity
violation terms $\mu^{'}_i L_i H_u$, the right-handed sneutrino will
decay into lepton pairs with a large fraction, due to mixing between
Higgs and left-handed lepton superfields. However, it should be
mentioned that mixing between right-handed sneutrino and Higgs boson
can not be avoided in general. These mixing terms have to be
suppressed in order not to conflict with PAMELA antiproton data. The
kinematical conditions can be used to suppress the right-handed
sneutrino decay to two Higgs or one Higgs plus one lepton. To
suppress the LSP direct decay to quarks which is induced by the
mixing of right-handed sneutrino and Higgs, an extra relation
$\sum_i y_{ij}\mu^{'}_i \sim 0$ is needed for the j-th generation
right-handed sneutrino which is the LSP. In the
Ref.\cite{Chen:2008dh} authors proposed similar idea, they assume
the LSP is the third generation right-handed sneutrino and only
$\mu^{'}_1\neq0$, then the small yukawa coupling $y_{13}$ make this
right-handed sneutrino as a good candidate of LDDM. The difference
between leptonically decaying neutralino and right-handed sneutrino
might be represented in the original positron spectrum. The former
one is softer than the latter, because the former one is three-body
decay while the latter one is two-body decay. It means that the
PAMELA implication on DM mass is different for these two scenarios.
With the synergy of PAMELA and collider experiments, it is possible
to distinguish whether the DM is neutralino, gravitino or
right-handed sneutrino.

\section{Summary and discussion}

The recent released PAMELA data shows interesting features,
that the positron shows an obvious excess above $\sim 10$ GeV
while the antiproton flux is well consistent with the
expectation from the conventional cosmic ray model.
The result implies there should exist some kinds
of primary positron sources in addition to the secondaries from
CRs interactions with the ISM.

Before the discussion of possible exotic sources of positrons it is
extremely important to explore the background carefully. In this
work we first recalculate the background contributions of positron
and antiprotons from CRs using GALPROP. Considering the CR data of
unstable secondaries, such as $^{10}$Be/$^{9}$Be, it is possible to
reduce the uncertainties in predicting the positron and antiproton
flux from the propagation parameters \cite{Delahaye:2008ua}. Two
typical propagation models, the diffusion+convection and
diffusion+reacceleration ones, are adopted as benchmarks of the CR
propagation models. In both of the two models, propagation
parameters are adjusted so that most of the CR spectra, such as B/C,
$^{10}$Be/$^{9}$Be and so on, are well consistent with the
observation. The DC model is found to be consistent with the
$\bar{p}/p$ and the low energy $e^+/(e^++e^-)$ data of PAMELA
without introducing a charge-dependent solar modulation model. While
the DR model shows a bit of underestimation of the $\bar{p}/p$ data
and produce more low energy positrons. In both models the positron
fraction at high energies shows obvious excess.

We then consider DM origin of the primary positrons. We compared the
positron and antiproton spectra with PAMELA data by assuming the
annihilation or decay products are mainly gauge bosons, quarks and
leptons respectively. We find the PAMELA data exclude quarks being
dominant final states, disfavor gauge bosons and favor the leptonic
final states. Comparing annihilating DM and decaying DM scenarios,
we prefer decaying DM because annihilating DM usually requires very
large boost factors. Moreover, annihilating DM usually produce more
gauge bosons and quarks than leptons.

A concrete example of such LDDM model is considered in MSSM with
tiny $ LL\bar E$ R-parity violation term. We choose neutralino as
the LSP and the DM particle. We show that neutralino with mass $600
\sim 2000$ GeV and life time of $\sim 10^{26}$ seconds can fit the
PAMELA data very well. We also demonstrate that neutralino with mass
around 2TeV can fit PAMELA and ATIC data simultaneously. Another
LDDM model is right-handed sneutrino being DM with bilinear R-parity
violation term. With the interplay of PAMELA and LHC it is possible
to distinguish the DM between neutralino and right-handed sneutrino.

Finally we will discuss the implications of the LDDM models. LDDM
can have many interesting phenomena in experiments other than
PAMELA, such as gamma ray observation. In general, it can have
two-body decay and three-body decay channels as $DM \to l^ +  l^ - $
and $ DM \to l^ +  l^ -  X$ where the leptons in the final-states
are required from PAMELA data. Particle X can be photons $ \gamma$,
neutrino $ \nu$ in general. If the LDDM has the decay channel $DM
\to l^ + l^ - \gamma$, the spectrum of gamma ray can reveal the
property of LDDM and the mechanics behind. For example, if there is
an intermediating resonance particle Y which decays to lepton pairs,
the decay of LDDM is now $ DM \to \gamma Y \to l^ +  l^ - \gamma$.
Thus the photon is monochromatic and can be explored by the Fermi
Gamma Ray Space Telescope(FGRST). Detecting the gamma ray spectrum
will be an important way to test the LDDM model, where FGRST will
play an important role.

For our neutralino LDDM model the neutrino observation will provide
a way to test the model since the decay produces hard neutrino
spectrum. Another possible way to test the model is to look for the
synchrotron emission of the hard electrons and positrons by DM
decay. Further studies on this issue are on going.

\acknowledgements

P.F.Y and J.L thank H. Murayama for useful discussion at Jiuhua,
Beijing. X.J.B thanks J. Chang for helpful discussions. This work
was supported in part by the Natural Sciences Foundation of China
(Nos. 10775001, 10635030, 10575111, 10773011), by the trans-century
fund of Chinese Ministry of Education, and by the Chinese Academy of
Sciences under the grant No. KJCX3-SYW-N2.

\end{document}